\documentclass[%
 aps,
 jmp,%
 amsmath,amssymb,
 twocolumn, 
 showpacs,
]{revtex4-1}

\usepackage{graphicx}% Include figure files
\usepackage{dcolumn}% Align table columns on decimal point
\usepackage{bm}% bold math
\usepackage{color}

\begin{document}

%\preprint{AIP/123-QED}

\title[Sample title]{
Metal-insulator transition caused by the coupling to \\
localized charge-frustrated systems under ice-rule local constraint
}

\author{Hiroaki Ishizuka}
\author{Masafumi Udagawa}
\author{Yukitoshi Motome}
\affiliation{ 
Department of Applied Physics, University of Tokyo, Hongo, 7-3-1, Bunkyo-ku, Tokyo 113-8656, Japan
}

\date{\today}

\begin{abstract}
We report the results of our theoretical and numerical study on electronic and transport properties 
of fermion systems with charge frustration.
We consider an extended Falicov-Kimball model in which itinerant spinless fermions interact repulsively
by $U$ with localized particles whose distribution satisfies a local constraint under geometrical frustration, 
the so-called ice rule. 
Electronic states of the itinerant fermions are studied by approximating the statistical average by the arithmetic mean 
over different configurations of localized particles under the constraint.
We numerically calculate the density of states, optical conductivity, and inverse participation ratio 
for the models on the pyrochlore, checkerboard, and kagome lattices, and discuss the nature of metal-insulator transitions at commensurate fillings. 
The results are compared with exact solutions for the models on Husimi cacti as well as with numerical results
for completely random distribution of localized particles.  
As a result, we show that the ice-rule local constraint leads to
several universal features in the electronic structure common to different lattice structures;
a charge gap opens at a considerably small $U$ compared to the bandwidth, and 
the energy spectrum approaches a characteristic form in the large $U$ limit, 
that is, the noninteracting tight-binding form in one dimension or the $\delta$-functional peak. 
In the large $U$ region, the itinerant fermions are confined in the macroscopically-degenerate ice-rule configurations, 
which consist of a bunch of one-dimensional loops: We call this insulating state the charge ice. 
On the other hand, transport properties are much affected by the geometry and dimensionality of lattices; 
e.g., the pyrochlore lattice model exhibits a transition from a metallic to the charge-ice insulating state 
by increasing $U$,
while the checkerboard lattice model appears to show Anderson localization before opening a gap.
Meanwhile, in the kagome lattice case, we do not obtain clear evidence of Anderson localization.
Our results elucidate the universality and diversity of phase transitions to the charge-ice insulator
in fully frustrated lattices.
\end{abstract}

\pacs{
71.10.Fd, % Lattice Fermion Models
71.27.+a, % Strongly correlated electron systems; heavy fermions
71.30.+h  % Metal-insulator transitions and other electronic transitions
}% PACS, the Physics and Astronomy
 % Classification Scheme.

\maketitle

\section{\label{sec:intro}
Introduction
}
Charge frustration has recently attracted considerable attention, 
since it underlies many intriguing phenomena such as electronic ferroelectricity,
multiferroicity, and 
novel metallic state in quantum melting of charge order~\cite{vdBrink2008,Ishihara2010,Seo2006}. 
In these systems, frustration often prevents the system from stabilizing a long-range order, 
and results in a ground state with macroscopic degeneracy. 
The degenerate manifold is highly sensitive to perturbations, 
such as couplings to other degrees of freedom, quantum and thermal fluctuations, 
and external stimuli such as application of pressure or external fields.
This is a source of fascinating phenomena in charge frustrated systems. 

In the case of localized electrons, the problem is well described by the Ising models on frustrated lattices,
in which spin up and down represent charge rich and poor states. 
A classic example is found in a charge ordering on the frustrated pyrochlore lattice, 
which was argued as the origin of the metal-insulator transition in magnetite Fe$_3$O$_4$~\cite{Verwey1939}. 
It was pointed out that the problem can be mapped onto an antiferromagnetic Ising model 
on the pyrochlore lattice, and the strong frustration precludes the nearest-neighbor Coulomb repulsion
to stabilize a charge ordering~\cite{Anderson1956}.  
The degenerate ground-state configurations consist of arbitrary network of locally-correlated tetrahedra,
in which every tetrahedron has two charge-rich and two charge-poor sites.
This local constraint is equivalent to the one discussed for proton configurations in water
ice~\cite{Bernal1933,Pauling1935}, and is called the ice rule.
More recently, a magnetic analogue, the spin ice, 
was discovered in several pyrochlore oxides~\cite{Harris1997,Ramirez1999},
which has been promoting the understanding of the ice-rule physics~\cite{Bramwell2005}. 

One of the most striking features of such ice-rule systems is 
cooperative nature of the spatial correlation. 
It was argued that the local constraint brings about a hidden gauge structure:
The ice rule corresponds to a zero-divergence condition 
in terms of a notional electric (or magnetic) field, which leads to 
a dipolar correlation in the ice-rule variables~\cite{Isakov2004,Henley2005}.
Systems under the local constraint are not simply disordered but have rather cooperative nature.

Recently, the effect of ice rule on electron itinerancy has also attracted increasing interest. 
It is highly nontrivial how the cooperative nature of the ice-rule degenerate manifold 
affects electronic and transport properties. 
Experimentally, anomalous transport phenomena were observed in several pyrochlore-based compounds 
and their relation to the ice-rule degeneracy has been discussed. 
For example, peculiar magneto-transport phenomena were found in 
hybrid compounds of itinerant $d$ electrons and localized $f$ moments 
in Nd$_2$Mo$_2$O$_7$~\cite{Taguchi2001} and Pr$_2$Ir$_2$O$_7$~\cite{Nakatsuji2006,Nakatsuji2010}, 
in which the $f$ moments potentially have spin-ice type correlations.
Spin-ice like frustration in charge and orbital degrees of freedom might also be relevant to 
a heavy-mass behavior and a metal-insulator transition under pressure in
a mixed-valence compound LiV$_2$O$_4$~\cite{Kondo1997,Urano2000,UranoPhD,Takeda2005,Pinsard-Gaudart2007}.
Theoretically, although several interesting aspects were pointed out, 
such as a fractional charge excitation~\cite{Fulde2002} and 
a lifting of ice-rule degeneracy by kinetic motion of electrons~\cite{Shimomura2004,Shimomura2005}, 
much less is known about the role of ice rule in itinerant systems, compared to that in localized systems. 

%{\bf The aim of this study is to elucidate the effects of localized systems with local constraints on
%the macroscopic properties of itinerant electrons, and not on explaining an experimentally observed
%phenomena.
The aim of this study is to elucidate
%In particular, we focus on
how the coupling to a localized system with the ice-rule local constraint
affects electronic and transport properties of itinerant electrons.
We try to reveal the fundamental and universal aspects of such systems, instead of explaining each
specific experimental data in real compounds.
We examine this problem by considering an extended Falicov-Kimball model on geometrically frustrated lattices, 
such as the pyrochlore, checkerboard, and kagome lattices.
The model describes itinerant spinless fermions interacting with localized particles whose spatial distribution obeys the ice rule. 
We numerically calculate electronic properties of the model by taking the arithmetic mean over different
ice-rule configurations of localized particles, instead of the statistical average. 
This is an approximation that becomes exact when different ice-rule configurations have the identical Boltzmann weight:
It is indeed the case of the models on Husimi cacti for which we obtain exact solutions for comparison.
We discuss to what extent the approximation holds in the other lattice systems. 

Our main result is the clarification of universality and diversity 
of the nature of metal-insulator transition caused by the coupling to the ice-rule manifold.
We find that the electronic structure exhibits several universal features irrespective of the lattice structures.
In particular, the systems commonly show a gap opening as increasing the repulsive interaction at
a commensurate filling. 
The gapped insulator in the strong coupling region exhibits a peculiar electronic state; 
itinerant fermions are localized in the specific ice-rule configuration --- we call it the charge-ice insulator. 
In contrast, transport properties are sensitively dependent on the lattice structure.
For the three-dimensional pyrochlore-lattice model, our numerical results indicate 
a direct phase transition from a metal to the charge-ice insulating state.
On the other hand, the two-dimensional checkerboard lattice appears to exhibit Anderson localization,
namely, a metal-to-insulator transition without gap opening, before going into the charge-ice insulator. 
For the kagome lattice, however, the results do not show any clear sign of Anderson localization.
We discuss in detail the universal and diverse aspects of the effect of ice rule on electron itinerancy 
through the comparison with the exact solutions for the cactus models
as well as the numerical results for completely random distribution of localized particles. 

The organization of this paper is as follows.
In Sec.~\ref{sec:model_and_method}, we introduce models and methods. 
After introducing the model Hamiltonian and the ice-rule limit in Sec.~\ref{sec:model}, 
we describe how we compute the electronic properties in Sec.~\ref{sec:averaging} and \ref{sec:numerics}. 
The validity of the approximation introduced in the calculation is examined in Appendix A. 
We present the various lattice structures in Sec.~\ref{sec:lattices}, and 
describe the exactly-solvable cases, Husimi cacti, in Sec.~\ref{sec:cactus}.
In Sec.~\ref{sec:result}, we show our results on electronic and transport properties of the 
models on the pyrochlore, checkerboard, and kagome lattices, 
in comparison with the exact solutions for Husimi cacti  
and the numerical results for uncorrelated random distributions. 
The results are discussed in detail in Sec.~\ref{sec:discussion}.  Sec.~\ref{sec:summary} is devoted to summary. 

\section{\label{sec:model_and_method}
Model and Method
}

\subsection{
Model
\label{sec:model}
}

We start with an extended Falicov-Kimball model~\cite{Udagawa2010} (ex-FK model) on the pyrochlore
lattice [Fig.~\ref{fig:lattice}(a)], 
\begin{eqnarray}
H= &-&t\sum_{\left<i,j\right>} (c_i^\dagger c_j + {\text{h.c.}}) \nonumber \\
 &+& U\sum_i n_i^c(n_i^f-\frac{1}{2})
 + V \sum_{\left<i,j\right>} n_i^f n_j^f, \label{eq:hamiltonian_fk}
\end{eqnarray}
where $c_i$ ($c_i^\dagger$) is an annihilation (creation) operator for an itinerant spinless fermion 
at site $i$ and $n_i^c$ is the number operator, $n_i^c = c_i^\dagger c_i$. 
$n_i^f$ denotes the number of localized classical particles at site $i$; $n_i^f = 0$ or $1$. 
The sum $\left<i,j\right>$ runs over the nearest-neighbor sites.
The first term describes the hopping of itinerant fermions,
the second term is the repulsive interaction between itinerant fermions and localized particles ($U>0$),
and the third term represents the nearest-neighbor interaction between localized particles.
We take an energy unit as $t=1$ except when explicitly shown. 

Hereafter we focus on the ``ice-rule limit": 
(i) the total number of localized particles is fixed at $N/2$, with $N$ 
      being the total number of sites,
and (ii) the interaction between localized particles is taken to be positive infinity, i.e., $V/t \rightarrow \infty$.
In this limit, the localized particles are distributed over the system with satisfying the ice-rule constraint;
two out of four sites are occupied by localized particles in every tetrahedron, 
as exemplified in Fig.~\ref{fig:lattice}(a). 
We call the model in the ice-rule limit the ice-rule model hereafter. 

Later, we extend the model (\ref{eq:hamiltonian_fk}) to other lattices, 
the checkerboard and kagome lattices in Sec.~\ref{sec:lattices}.
We also consider the models on the Husimi cactiin Sec.~\ref{sec:cactus},
for which the analytical solutions are available. 
We will introduce a similar ice-rule appropriately in each case. 

\subsection{
Arithmetic mean approximation within the ice-rule manifold \label{sec:averaging}
}

For the model (\ref{eq:hamiltonian_fk}), the expectation value of an observable $\hat{A}$ is given by
\begin{eqnarray}
\langle\hat{A}\rangle= \frac{{\rm Tr}_f {\rm Tr}_c\hat{A}\exp(-\beta H)}{{\rm Tr}_f {\rm Tr}_c \exp(-\beta H)},
\end{eqnarray}
where ${\rm Tr}_c$ (${\rm Tr}_f$) is the trace over configurations of itinerant fermions (localized particles)
and $\beta$ is the inverse temperature. 
For a given configuration $\{n_i^f\}$, the Hamiltonian (\ref{eq:hamiltonian_fk}) is reduced to 
a one-body Hamiltonian with onsite potential: 
\begin{eqnarray}
H(\{n_i^f\}) =  -t\sum_{\left<i,j\right>} (c_i^\dagger c_j + {\text{h.c.}}) + \sum_i U_i n_i^c, \label{eq:hamiltonian_onsite}
\end{eqnarray}
where the binary potential is given by
\begin{eqnarray}
U_i= U (n_i^f-\frac12 ) = +\frac{U}{2} \ \ \text{or} \ -\frac{U}{2},  
\end{eqnarray}
corresponding to $n_i^f = 1$ or $0$. 
Since we consider the ice-rule limit here, the spatial distribution of $U_i$ obeys the ice-rule constraint;
two sites of every tetrahedron being $+U/2$ and the other two being $-U/2$.
Then, the expectation value $\langle\hat{A}\rangle$ is rewritten as
\begin{eqnarray}
\langle\hat{A}\rangle= \frac{\sum_{\{ n_i^f\} \in \sf{ice}} \langle 
\psi_g 
|\hat{A}| 
\psi_g 
\rangle
\exp(-\beta 
E_g )}{\sum_{\{ n_i^f\} \in \sf{ice}}\exp(-\beta 
E_g )},
\label{observable}
\end{eqnarray}
where $E_g = E_g(\{n_i^f\})$ and $|\psi_g\rangle = |\psi_g(\{n_i^f\})\rangle$ 
are the energy and many-body eigenfunction of the ground state of
$H ( \{n_i^f\} )$; 
$H ( \{n_i^f\} ) |\psi_g(\{n_i^f\})\rangle = E_g(\{n_i^f\}) |\psi_g(\{n_i^f\})\rangle$. 
Here, we assume 
the temperature to be sufficiently low, and ignore the contribution from excited states.
The sum over $\{ n_i^f \} \in \sf{ice}$ is taken for all configurations of the binary
onsite potential $U_i$ that obey the ice rule.

Then, by computing the eigenvalues and eigenstates for all the ice-rule configurations, 
one obtains the statistical average of observables. 
However, it is virtually impossible to take the sum in Eq.~(\ref{observable}) 
since the number of ice-rule configurations
grows exponentially with increasing the system size ($\sim 1.5^{N/2}$)~\cite{Pauling1935,Nagle1966}. 
To avoid this difficulty and to extract the essential physics of itinerant electrons 
coupled with the ice-rule localized variables, 
we replace the statistical average by the arithmetic mean
with omitting the Boltzmann weight in Eq.~(\ref{observable}). 
This corresponds to the assumption 
that $E_g(\{n_i^f\})$ does not depend on 
different ice-rule configurations of onsite potential $U_i$, 
i.e., $E_g(\{n_i^f\}) = E_0$. 
Moreover, we calculate the average by sampling the ice-rule configurations randomly. 
Namely, we approximate Eq.~(\ref{observable}) by 
\begin{eqnarray}
\langle\hat{A}\rangle&\simeq& 
\frac{\sum_{\{ n_i^f\} \in \sf{ice}} \langle 
\psi_g |\hat{A}| 
\psi_g \rangle
\exp(-\beta E_0)}{\sum_{\{ n_i^f\} \in \sf{ice}}\exp(-\beta E_0)}\\
&=&
\frac{1}{N_{\rm{ice}}}\sum_{i=1}^{N_{\rm{ice}}} \sum_{m=1}^{N_c} \langle m |\hat{A}| m \rangle 
\label{simpleaverage}
\\
&\simeq&
\frac{1}{N_{\rm{samp}}}\sum_{i=1}^{N_{\rm{samp}}} \sum_{m=1}^{N_c} \langle m |\hat{A}| m \rangle,
\label{sampling}
\end{eqnarray}
where  $N_{\rm{ice}} \sim 1.5^{N/2}$ is the total number of ice-rule configurations of localized particles, 
$N_{\rm{samp}}$ is a number of samples in actual calculations,
and $N_c$ is the total number of itinerant fermions.
Here, $|m\rangle$ are the one-particle eigenstates of $H ( \{n_i^f\} )$, and are sorted in ascending order of the eigenenergies. 

The arithmetic mean in Eq.~(\ref{simpleaverage}) becomes exact 
if the Boltzmann weight in Eq.~(\ref{observable}) 
is identical for all different ice-rule configurations of localized particles in the ground state. 
This is indeed the case for models on Husimi cacti~\cite{Udagawa2010} as discussed in Sec.~\ref{sec:cactus}: 
In these models, because of the loopless structure of lattices, 
the ice-rule configurations are all topologically equivalent, 
which results in the identical Boltzmann weight. 
For the pyrochlore lattice case, however, the Boltzmann weights are not identical for different configurations; 
nevertheless, the differences turn out to be very small because of the structure of degenerate ice-rule manifold. 
In fact, we find that the difference of the total energy is typically in the order of $10^{-4} t$
in the entire region of $U/t$.
Detailed discussions will be given in Appendix~\ref{subsec:groundstate}. 
This suggests that the approximation in Eq.~(\ref{simpleaverage}) gives quantitatively reasonable results in the 
situation that we are interested in, where the ice-rule manifold is well
preserved and the system takes all the ice-rule states without selecting a unique ground state or a submanifold if any. 
In other words, the 
arithmetic mean provides a tractable and reasonable tool 
to investigate the effect of the ice-rule degeneracy on the electronic state of itinerant fermions. 

Through the procedure above, the many-body problem in Eq.~(\ref{eq:hamiltonian_fk}) is reduced 
to a one-body problem with onsite ice-rule potential in Eq.~(\ref{eq:hamiltonian_onsite}). 
For a reference to the ice-rule model, we also consider a one-body model with random potential. 
In this case, we consider completely random configurations of $N/2$ localized particles 
without any spatial correlation in the Hamiltonian (\ref{eq:hamiltonian_onsite}). 
The comparative study illuminates the effect of local correlation in the ice-rule model, 
as we will see later. 

In general, a special configuration of the potential may lead to a characteristic electronic state. 
For example, it was pointed out that, even in the one-dimensional case, 
a locally-correlated potential can drive the system delocalized~\cite{Dunlap1990}. 
In the following, we elucidate the effect of local correlation brought by the ice rule 
in higher dimensions, in comparison with the random cases without any spatial correlation. 

\subsection{
Numerical calculations
\label{sec:numerics}
}

We calculate the electronic properties of model (\ref{eq:hamiltonian_fk})
numerically by using Eq.~(\ref{sampling}). 
In the calculation, we need to generate $N_{\rm{samp}}$ samples of ice-rule potential configurations 
and obtain the eigenstates for each sample. 
To generate different ice-rule samples sequentially,
we employ the so-called loop algorithm~\cite{Rahman1972,Barkema1998}.
To retain the statistical independence, we applied $10^6$ loop updates between the samples. 
This is much larger than the ``autocorrelation time" $\tau$, typically in the order of tens~\cite{autocorr}.
On the other hand, we also examined the convergence of Eq.~(\ref{sampling}) as to $N_{\rm{samp}}$. 
We checked $N_{\rm{samp}}$ dependence and conclude that average over $N_{\rm{samp}} = 40$ gives converged results
with enough precision for the following discussions. 
Statistical errors are fairly small for the density of states (DOS), 
which are comparable to the width of curves in the plots below. 
For other quantities, the errors are explicitly shown in the plots. 
Once the sample set is obtained, we calculate the eigenenergies and eigenstates 
by the exact diagonalization of the Hamiltonian given by Eq.~(\ref{eq:hamiltonian_onsite}). 

In the following sections, we discuss the electronic and transport properties by calculating 
DOS and the optical conductivity.  
DOS is obtained from the eigenenergies by taking the histogram over $N_{\rm{samp}}$ samples 
with an energy width $\Delta \varepsilon= 0.02$. 
The energy gap is directly calculated by the arithmetic mean of the energy differences 
between the highest occupied level and the lowest unoccupied level. 
The optical conductivity is calculated by the standard Kubo formula:
 
\begin{eqnarray}
\sigma(\omega) = \sum_{m \neq n}^{N_c} 
\frac{f(\varepsilon_n)-f(\varepsilon_m)}{\varepsilon_m - \varepsilon_n}
\big|\langle m | {\bf J}_\mu | n \rangle \big|^2
\delta( |\varepsilon_m - \varepsilon_n| - \omega ),\nonumber \\ \label{eq:kubo}
\end{eqnarray}
where $f(\varepsilon)$ is the Fermi distribution function and
\begin{eqnarray}
{\bf J}_\mu = -i t\sum_{\langle j,k \rangle} ( {\bf n}_\mu \cdot {\boldsymbol \delta}_{j,k} )( c_k^\dagger c_j - c_j^\dagger c_k )
\end{eqnarray}
is a current operator in the $\mu$ direction ($\mu$ is assigned for each case below),
which is constructed in a standard way from a polarization operator in order to
satisfy the continuity equation~\cite{Mahan2000}.
Here, ${\bf n}_\mu$ is the unit operator in the $\mu$ direction, $t$ is the
transfer integral for the nearest-neighbor sites, and
$\boldsymbol \delta_{j,k}$ is the geometrical vector from $j$th to $k$th site.
The sum is taken for all the nearest-neighbor pairs.
To calculate the optical conductivity at $T\rightarrow 0$, we set $\beta = 10^6$.
As a measure of the metallicity, we also calculate the
low-energy weight of $\sigma(\omega)$ defined by
\begin{eqnarray}
w = \int^{\omega_0}_{0}  \sigma(\omega) 
\, d\omega
\label{eq:neff}
\end{eqnarray}
with $\omega_0$ being a cut off.

To further examine the metallicity of the system, 
in particular, to detect Anderson localization, 
we also compute the inverse participation ratio (IPR)~\cite{Kramer1993}.
IPR is defined by 
\begin{eqnarray}
P^{-1}=\sum_l |\phi({\mathbf R}_l)|^4,
\label{eq:IPR}
\end{eqnarray}
where $\phi({\mathbf R_l})$ denotes an eigenfunction at site ${\mathbf R}_l$. 
IPR extrapolated to the infinite system size $N$ gives a measure of the localization of the eigenfunction: 
If $\phi$ is an extended state, IPR behaves as $ P^{-1} \propto N^{-1} $
$\rightarrow 0$ 
when $N\rightarrow \infty$, while if $\phi$ is localized, $P^{-1}$ 
converges to a nonzero value as $N\rightarrow \infty$. 
Usually, IPR is calculated for the state right at the Fermi level, i.e., for the highest occupied state. 
In the following calculations, however, we focus on a special filling at which
DOS shows divergence at the Fermi level due to the degeneracy of a huge number of states. 
(See, e.g., Fig.~\ref{fig:pyro:dos}.) In such cases, IPR 
right at the Fermi level is not well defined because of the degeneracy. 
Hence we measure the metallicity by calculating IPR for the states just above the flat bands.
The evaluation is conducted by the harmonic average of IPR for the states within a small energy window
with the width of $\Delta \varepsilon = 0.02$, so that the averaged IPR goes to zero as $N\rightarrow\infty$
if the states in the energy window include at least one extended state~\cite{IPR}.

\subsection{
Lattices
\label{sec:lattices}
}

\begin{figure}
  \begin{center}
    \includegraphics[width=2.95in]{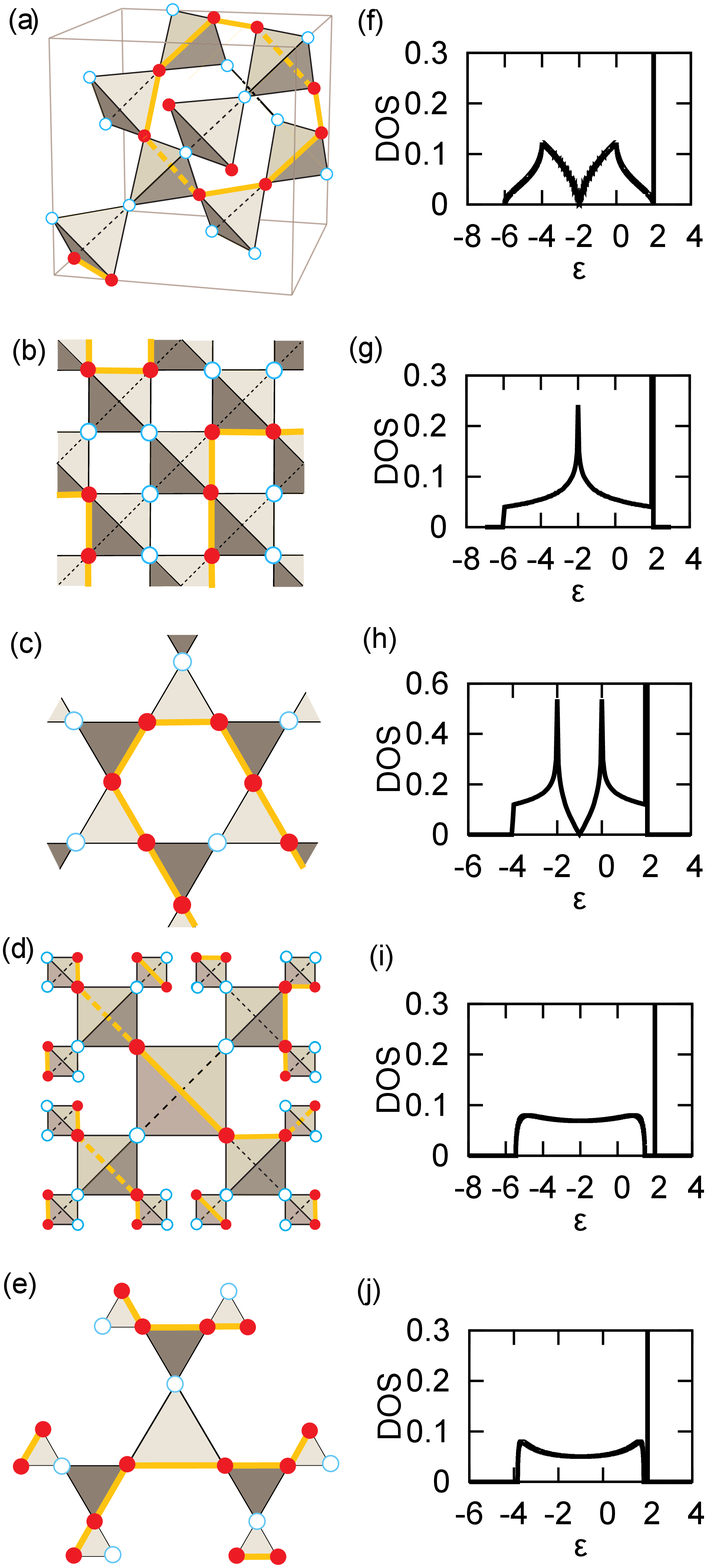}
  \end{center}
  \caption{
  (color online).
  Lattice structures of 
  (a) pyrochlore, (b) checkerboard, (c) kagome lattices, and (d) tetrahedron 
  and (e) triangular Husimi cacti. 
  DOS for the noninteracting tight-binding model on each lattice structure are shown in (f)-(j).
  In the lattice figures, filled (open) circles denote the sites unoccupied (occupied) by localized particles. 
  The configurations represent examples of the ice-rule configurations.  
  Loops connecting the unoccupied sites are shown by the bold lines. 
  See text for the details.
  }
  \label{fig:lattice}
\end{figure}

The model (\ref{eq:hamiltonian_fk}) is introduced for the pyrochlore lattice shown in Fig.~\ref{fig:lattice}(a). 
In the following, we also consider similar models on the checkerboard and kagome lattices, 
as shown in Figs.~\ref{fig:lattice}(b) and \ref{fig:lattice}(c). 
These lattice structures are two-dimensional cousins of the three-dimensional pyrochlore lattice; 
the checkerboard lattice is a $\langle 001\rangle$ projection of the pyrochlore lattice, 
while the kagome is a $\langle 111\rangle$ plane. 
The checkerboard lattice shares its geometrical unit, a tetrahedron, with the pyrochlore lattice, 
while the unit of the kagome lattice is a triangle. 
On the other hand, the kagome lattice shares a global geometrical feature with the pyrochlore lattice, 
i.e., the smallest loop composed of the geometrical units is a hexagon 
as in the pyrochlore case; 
whereas it is a square in the checkerboard case. 
Through the comparative study among these different lattices, 
we examine the effect of differences in dimensionality, geometrical unit, and their connection. 

For the checkerboard lattice,
the ice-rule local constraint is applied in the same manner as for the pyrochlore lattice.
Namely, we consider $N/2$ localized particles, and place two of them on each tetrahedron,
as in Fig.~\ref{fig:lattice}(b). 
On the other hand, for the kagome lattice, we consider an ice-rule type constraint by distributing $N/3$
localized particles so that one site is occupied and the other two are unoccupied in each triangle, 
as exemplified in Fig.~\ref{fig:lattice}(c).  We call this rule the ``kagome ice rule" hereafter.

All three lattice structures share common features; 
they consist of corner sharing network of geometrically-frustrated units, tetrahedra or triangles. 
These units are called the complete graph, i.e., a graph in which all vertices are 
connected with all the other vertices. 
To illuminate the role of the common features, we also consider a set of variants for these lattices, 
that is, the so-called Husimi cacti of the complete graphs, as discussed separately in the next section.  

\subsection{
Exactly solvable models on the Husimi cacti
\label{sec:cactus}
}

In addition to the lattice structures introduced in the previous sections, 
we also consider modified structures composed of the same geometrical units 
--- the Husimi cacti of tetrahedra and triangles.
They are the analogues of the Bethe lattice composed of corner-sharing tetrahedra or triangles, 
which we call here the tetrahedron or triangle Husimi cactus, respectively 
[Figs.~\ref{fig:lattice}(d) and \ref{fig:lattice}(e)].
The Husimi cacti share two important structural features with the original lattices; 
the geometrical units and their corner-sharing network. 
A difference is in the global connection of the units: As mentioned in Sec.~\ref{sec:lattices}, 
all the three lattices have loops running across different  geometrical units, 
but the Husimi cacti do not have such global loops. 
Thus, the Husimi cacti are the loopless variants of the original lattices. 

A crucial advantage of considering the Husimi cacti is that the ex-FK model in the ice-rule
limit is exactly solvable. 
The benefit is a direct consequence of the loopless structure, 
which makes different ice-rule configurations topologically equivalent. 
The derivation of the exact solution for the tetrahedron Husimi cactus was already reported in
Ref.~\cite{Udagawa2010}, and it is straightforward to extend the method to the triangle Husimi cactus case. 
Hence, here we only show the final results.  For the cacti models, the local Green's function at site $i$, 
$G_{i}(\varepsilon) \equiv t 
\langle i|
[\varepsilon - 
\mathcal{H} (\{ n^f_i \}) +i\delta]^{-1} |i\rangle$, depends not explicitly on $i$ but only on the value of $U_i$;
i.e.,  it is possible to write $G_i = G_{\pm}$ corresponding to the sites with $U_i = \pm U/2$. 
(We include $t$ in the definition of $G$ to simplify the following expressions.)
$G_{\pm}$ are obtained as
\begin{eqnarray}
G_\pm^{-1} = \frac{2}{g_\pm} - \frac{1}{t}\big(\varepsilon \mp \frac{U}{2} \big),
\label{fullDyson}
\end{eqnarray}
where $g_{\pm}$ are the solutions of the following recursive equations for each case:
(i) for the tetrahedron Husimi cactus with $\sum_i n^f_i = N/2$,
\begin{eqnarray}
\frac{g_\pm(1-g_\mp)+2g_\mp(1-g_\pm)}{1+g_\mp(1-2g_\pm)} 
= \frac{1}{t}\big( \varepsilon \mp \frac{U}{2} \big) - \frac{1}{g_\pm},
\label{g_tetrahedron}
\end{eqnarray}
%
\if0{
\item triangle Husimi cactus with $\sum_i n^f_i = 2N/3$
\begin{eqnarray}
\left\{\begin{array}{l}
\displaystyle{\frac{g_+ + g_- - 2g_+g_-}{g_+g_- - 1}
= \frac{1}{t}\big( \varepsilon - \frac{U}{2} \big) - \frac{1}{g_+}},\\
%
\displaystyle{\frac{2g_+}{g_+ + 1}
= \frac{1}{t}\big( \varepsilon + \frac{U}{2} \big) - \frac{1}{g_-}},
\end{array}\right.
\label{g_triangle21}
\end{eqnarray}
}\fi
and (ii) for the triangle Husimi cactus with $\sum_i n^f_i = N/3$,
\begin{eqnarray}
\left\{\begin{array}{l}
\displaystyle{\frac{2g_-}{g_- + 1}
= \frac{1}{t}\big( \varepsilon - \frac{U}{2} \big) - \frac{1}{g_+}},\\
\displaystyle{\frac{g_+ + g_- - 2g_+g_-}{g_+g_- - 1}
= \frac{1}{t}\big( \varepsilon + \frac{U}{2} \big) - \frac{1}{g_-}}.
\end{array}\right.
\label{g_triangle12}
\end{eqnarray}

Equation~(\ref{fullDyson}) with the solutions of Eqs.~(\ref{g_tetrahedron}) and (\ref{g_triangle12}) 
gives the exact local Green's functions of the ex-FK models given by Eq.~(\ref{eq:hamiltonian_fk})
on the Husimi cacti in the ice-rule limit~\cite{Udagawa2010, Udagawa2010_2}. 
As we will show in the following sections, 
the solutions for Husimi cacti models give good references to the original lattice models. 

\section{\label{sec:result}
Results
}
In this section, we discuss the effect of ice-rule constraint on  
electronic and transport properties of 
models on different lattice structures one by one.
Comparisons with the results for models with random potential 
and the exact solutions for the Husimi cacti 
are also given.

%
%=============================================================================%
% RESULT SECTIONS:                                                            %
%                                      

\subsection{\label{subsec:pyro}
Pyrochlore lattice
}
\begin{figure*}
  \begin{center}
  \includegraphics[width=4.27in]{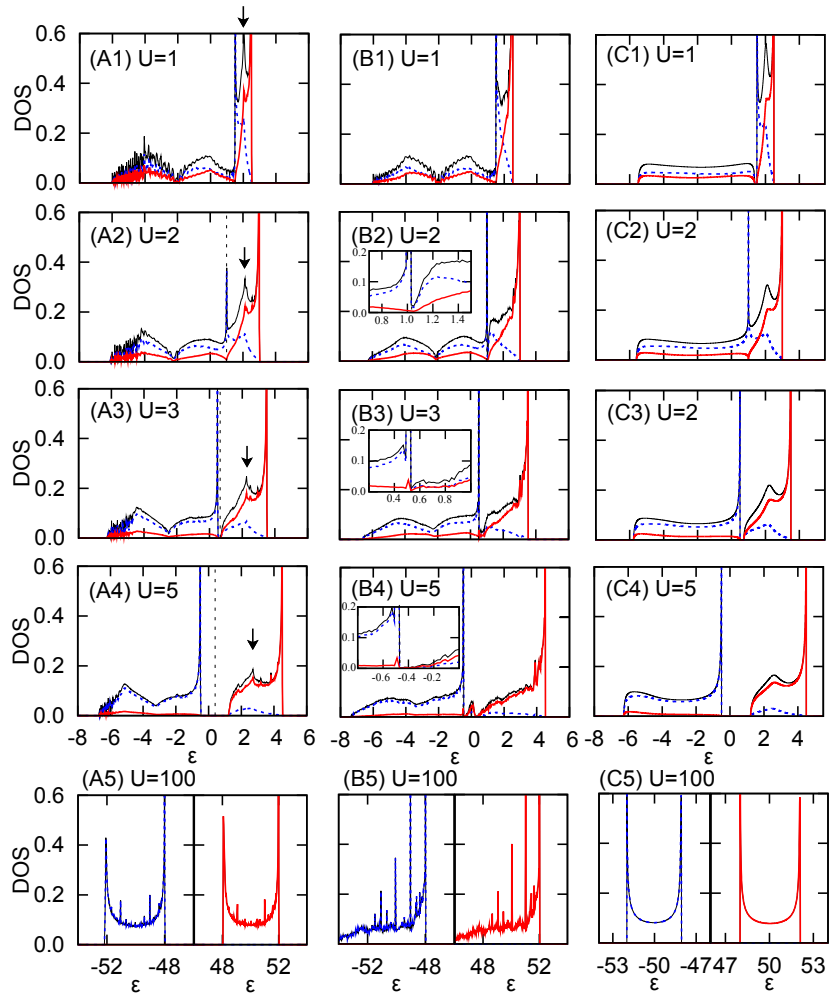}
  \end{center}
  \caption{
  (color online).
  DOS of itinerant fermions for the pyrochlore lattice models with ice-rule constraint (A1)-(A5)
  and with the random potential (B1)-(B5);
  DOS for the tetrahedron Husimi cactus model with the ice-rule constraint (C1)-(C5).
  The insets of (B2), (B3), and (B4) show the enlarged figures of the main panels 
  in the vicinity of the Fermi level at half filling.
  Bold (dotted) curves represent the partial DOS at the sites with potential $+U/2$ ($-U/2$).
  Thin curves represent the total DOS.
  The Fermi level for half-filling case is indicated by the vertical dashed lines.
  Error bars are within the width of the curves.
  Arrows in the ice-rule cases indicate the cusp-like structures discussed in the text.
  }
  \label{fig:pyro:dos}
\end{figure*}

Figure~\ref{fig:pyro:dos} shows DOS for the pyrochlore lattice case. 
The left column (A1)-(A5) shows the results for the ice-rule models, 
the middle (B1)-(B5) for the random-potential, and the right (C1)-(C5) for the exact results for the tetrahedron Husimi cactus.
Different rows correspond to the data at different values of $U$. 
The bold (dotted) curves represent the site-resolved DOS, $\rho_+$ ($ \rho_-$) for $U_i = +U/2$ ($-U/2$) sites,
and the thin curves denote the total DOS $\rho = \rho_+ + \rho_-$.
The numerical results for the pyrochlore lattice (left and middle columns) are obtained for $3^3$ superlattices
of $4\times 8^3$  sites.
The results for the tetrahedron Husimi cactus (right column) are calculated from 
Eqs.~(\ref{fullDyson}) and (\ref{g_tetrahedron}).

In the pyrochlore case, at $U=0$, the energy levels consist of two flat bands and two dispersive bands
[Fig.~\ref{fig:lattice}(f)].
The flat bands give the $\delta$-functional peak at $\varepsilon=2$.
Meanwhile, the dispersive bands are equivalent to those of the diamond lattice.
This equivalence can be understood as a result of line-graph correspondence.
The dispersive bands form a continuum spectrum for $-6\leq \varepsilon \leq 2$ with two semimetallic dips;
one is at $\varepsilon = -2$, where two dispersive bands touch with each other, 
and the other is at $\varepsilon=2$, where the higher dispersive band touches the flat bands. 
At half-filling $\sum_i \langle n_i^c \rangle = N/2$, 
the Fermi energy is located at $\varepsilon=2$, right at the latter semimetallic point; 
i.e., two dispersive bands are fully occupied and the flat bands are empty. 

By switching on $U$, the flat bands are perturbed to be broadened, resulting in a spectrum sandwiched by  
two divergences in DOS, as shown in Figs.~\ref{fig:pyro:dos}(A1) and \ref{fig:pyro:dos}(B1).
The Fermi level at half filling is pinned at the lower-edge divergence for both the ice-rule and random cases. 
It is worthy to note that, in the ice-rule case, a cusp-like structure appears between the two divergences,
as indicated by arrows in Figs.~\ref{fig:pyro:dos}(A1)-(A4).  We return to this point below. 

\begin{figure}
  \begin{center}
  \includegraphics[width=2.80in]{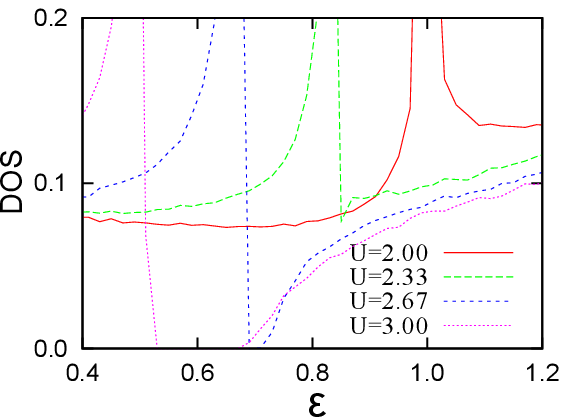}
  \end{center}
  \caption{
  (color online).
  DOS for the ice-rule model on the pyrochlore lattice in the vicinity of the Fermi level at half filling.
  }
  \label{fig:pyro:gap}
\end{figure}

As $U$ increases further, an energy gap starts to open both for the ice-rule and random cases.
However, the critical values of gap opening, $U_c$, are largely different between these two cases.
In the results for ice-rule case, there is a clear gap with $U\ge 3$
[Figs.~\ref{fig:pyro:dos}(A3) and \ref{fig:pyro:dos}(A4)],
whereas there still remains a small DOS at the Fermi level at half filling in the random case until $U\sim5$
[Figs.~\ref{fig:pyro:dos}(B3) and \ref{fig:pyro:dos}(B4)]. 
The behavior of the gap opening near $U_c$ is shown for the ice-rule model 
in Fig.~\ref{fig:pyro:gap}. 
$U$ dependence of the energy gap at half filling is summarized in Fig.~\ref{fig:pyro:list}(a). 
For the random case, a gap appears at $U \sim 6$, i.e.,
when the potential value becomes comparable to the bandwidth. 
On the other hand, the critical value of $U$ in the ice-rule case is estimated as $U_c = 2.3(2)$. 
Such a small $U_c$ compared to the bandwidth is 
considered to be characteristic of the correlated ice-rule configurations. 

In the ice-rule case for $U>U_c$, the system is insulating at half filling, where 
itinerant fermions are excluded from the sites with $U_i = +U/2$ and 
localized in the ice-rule configurations at $U_i=-U/2$ sites. 
We call this insulating state the charge-ice insulator~\cite{Udagawa2010}.
In the large $U$ limit, 
DOS for the charge-ice insulator approaches 
a pair of DOS for the one-dimensional (1D) tight-binding model
centered at $\varepsilon = \pm U/2$, as shown in Fig.~\ref{fig:pyro:dos}(A5).
This is in sharp contrast to 
the asymmetric featureless DOS for the random case in Fig.~\ref{fig:pyro:dos}(B5).
The upper-edge divergences of the two 1D-like bands at $\varepsilon \simeq \pm U/2 + 2$ come from
the two divergences in the perturbed flat bands in the small $U$ region described above. 
On the other hand, the lower-edge divergence in the upper band at $\varepsilon \simeq U/2 - 2$ develops
from the cusp-like feature seen already at $\varepsilon \simeq 2$ in the small $U$ region. 
The characteristic 1D-like feature is explained by the fact that, in the ice-rule case, 
the sites with $U_i = +U/2$ and $-U/2$ form 1D loops of equivalent onsite potential [see Fig.~\ref{fig:lattice}(a)].
The 1D loops have many different lengths from the shortest six site to infinite length; 
the average over the lengths for different ice-rule configurations results in the 1D-like DOS. 
The form of DOS shows slight deviations from that for the 1D tight-binding model; 
a dicernible feature is the spikes in the spectrum, which is presumably due to the finite length of loops.
Thus, the gapped insulating state at large $U$ is 
the charge-ice insulator in which itinerant fermions are confined in the 1D loops.

The interesting evolution of DOS with increasing $U$ for the ice-rule case 
is well reproduced by the tetrahedron Husimi cactus model [Figs.~\ref{fig:pyro:dos}(C1)-(C5)]~\cite{Udagawa2010}.
This model shows a transition to the charge-ice insulator at $U_c=2$, much smaller than the bandwidth,
similarly to the pyrochlore lattice model.
Furthermore, in the large $U$ limit, DOS converges to exactly the same form for 1D tight-binding model
centered at $\pm U/2$. 
This comes from the fact that the cactus lacks finite-length loops which 
are present in 
the pyrochlore lattice [Fig.~\ref{fig:lattice}(d)]. 
The cactus also well reproduces other important features, 
such as the overall form of DOS at $U=0$ (dispersive part plus flat bands~\cite{Udagawa2010}), 
[Fig.~\ref{fig:lattice}(i)], 
the split of the flat bands by $U$, and the cusp-like feature.
The agreement indicates that these universal features are 
owing to the peculiar corner-sharing geometry of tetrahedral units.

%
% Optical Conductivity
%

\begin{figure}
  \begin{center}
    \includegraphics[width=3.33in]{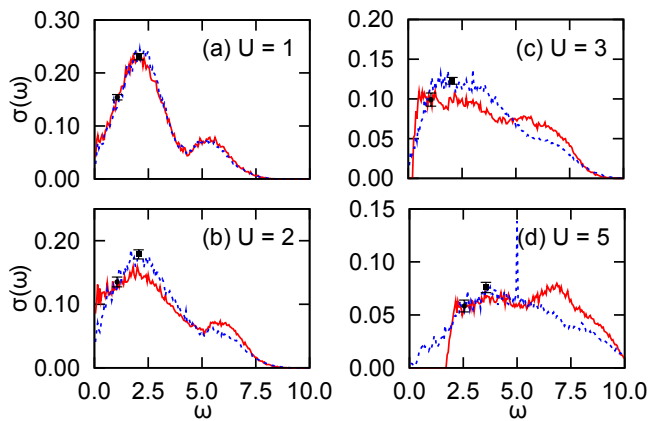}
  \end{center}
  \caption{
(color online).
  Optical conductivity calculated by Eq.~(\ref{eq:kubo}) for the pyrochlore lattice models at half-filling. 
  Solid (dotted) curves represent the results for the ice-rule
  (random) case. Typical errorbars are shown in each figure.
  }
  \label{fig:pyro:sigma}
\end{figure}

Now we turn to the transport properties. 
Figure~\ref{fig:pyro:sigma} shows the optical conductivity at half filling 
calculated along the [111] direction for $4^3$ superlattices of $4 \times 4^3$ sites.
The solid (dotted) curves show the results for the ice-rule (random) case.
As $U$ increases, $\sigma(\omega)$ evolves very differently for the ice-rule and random cases. 
In the random case, $\sigma(\omega)$ changes slowly with developing a dip at $\omega \sim 0$. 
In contrast, in the ice-rule case, the change is more drastic. 
First of all, a gap opens for $U>U_c$, consistent with the DOS in 
Figs.~\ref{fig:pyro:dos}(A3) and \ref{fig:pyro:dos}(A4). 
Furthermore, the low-energy part of $\sigma(\omega)$ does not decrease in a monotonic way before the gap opens; 
Comparing Figs.~\ref{fig:pyro:sigma}(a) and \ref{fig:pyro:sigma}(b),
we observe an increase of the low-energy part as $U \to U_c$.

This enhancement is more clearly seen in the
low-energy weight $w$ defined in Eq.~(\ref{eq:neff}).
The results are plotted in Fig. \ref{fig:pyro:list}(b).  Here we take the cutoff $\omega_0 = 0.095$; 
the results are qualitatively independent of $\omega_0$ when it is small enough.
The data clearly indicate the increase of $n_{\rm eff}$ with $U$ for the ice-rule case,
showing a maximum at $U\sim2$, and sharply drops at $U \sim U_c$,
in contrast to the featureless gradual decrease in the random case.
This characteristic behavior might be due to the peculiar semimetallic behavior at $U=0$; 
the metallicity is suppressed in the small $U$ region 
since there are less low energy states available due to the semimetallic dip. 
We will discuss this behavior in comparison with the result for the checkerboard lattice case 
in the next section.

Another peculiar difference in $\sigma(\omega)$ between the ice-rule and random cases is seen
in the large $U$ region, that is, a sharp spike only existing in the random case. 
In Fig.~\ref{fig:pyro:sigma}(d),
the data at $U=5$ exhibit a spike at $\varepsilon \simeq 5$ for the random case:
The spike comes from the transition between the divergences of DOS at $\varepsilon= \pm U/2 + 2$.
However, the spike is suppressed in the ice-rule case. 

The suppression is a consequence of quantum phase interference among the eigenstates at the divergence of DOS. 
The divergences  come from a particular set of states with 1D character, 
which are originally included in the flat bands at $U=0$.
In the ice-rule case, each wave function is confined in a 1D potential loop, and has
a uniform amplitude and an alternating sign as $+-+-+-\cdots$ along the loop. 
According to Eq.~(\ref{eq:kubo}), the contribution to $\sigma(\omega)$ from the transition 
between the two divergences is proportional to $\sum_{i,j} |\langle l_i|{\bf J}|h_j\rangle|^2$,
where $| l_i \rangle$ ($| h_i \rangle$) is an eigenstate in the lower (higher) divergence. 
Since the current operator ${\mathbf J}$ has a matrix element only between neighboring sites, 
the summation counts the matrix elements for a pair ($l_i, h_j$) 
for which the corresponding 1D loops have neighboring sites. 
There are, however,  cancellations between the matrix elements 
due to the alternating sign of the wave functions. 
This interference is the reason why the sharp spike is suppressed in the ice-rule case.

\begin{figure}
  \begin{center}
    \includegraphics[width=3.33in]{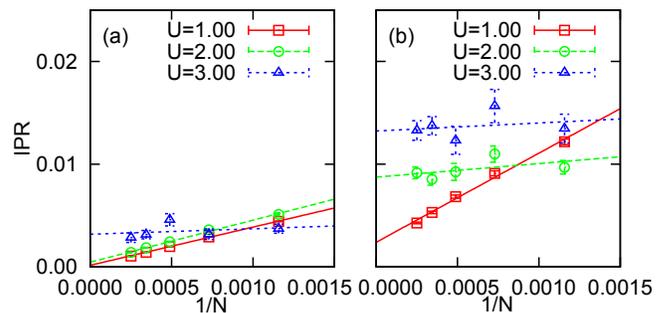}
  \end{center}
  \caption{
  (color online).
  IPR obtained by Eq.~(\ref{eq:IPR}) for the half-filled state in the pyrochlore lattice models 
  with (a) the ice rule and (b) random potential. The lines show the extrapolation to $N \to \infty$.
  See the texts for details.
  }
  \label{fig:pyro:ipr}
\end{figure}

\begin{figure}
  \begin{center}
  \includegraphics[width=2.67in]{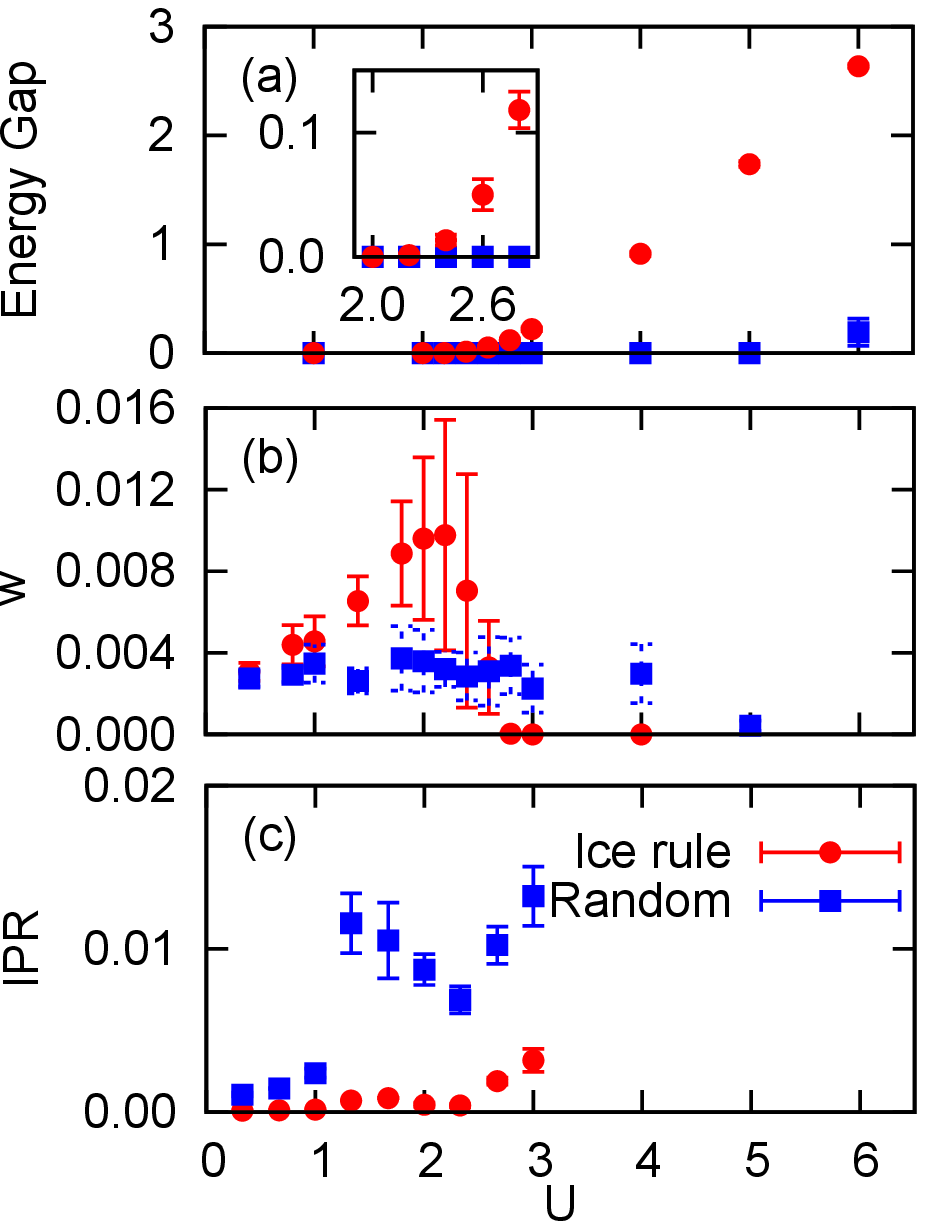}
  \end{center}
  \caption{
  (color online).
  $U$ dependence of (a) the energy gap, (b) the %effective carrier number
  low-energy weight of $\sigma(\omega)$, $w$%$n_{\text{LEW}}$
  , and
  (c) IPR for the pyrochlore lattice models at half filling.
  $w$ %$n_{\text{LEW}}$
  is calculated by Eq.~(\ref{eq:neff}) with taking the cutoff $\omega_0=0.095$, and
  IPR is estimated by the extrapolation to $N\rightarrow \infty$ 
  in Fig.~\ref{fig:pyro:ipr}. 
  The inset of (a) shows an enlarged view in the vicinity of $U_c = 2.3(2)$.
  }
  \label{fig:pyro:list}
\end{figure}

Difference in the nature of metal-insulator transition is also examined by using
the harmonic average of IPR, defined in the previous section. 
Figure~\ref{fig:pyro:ipr} shows the results. 
In the ice-rule case, IPR is extrapolated to zero as $N \rightarrow \infty$ for $U<U_c$, and becomes finite for $U>U_c$. 
This suggests that the system remains metallic until $U=U_c$ and the localization occurs simultaneously with the gap opening.
On the other hand, the extrapolated value in the random case becomes nonzero even at $U=1$.
This result indicates that the system is insulating because of the Anderson localization~\cite{Kramer1993},
since DOS remains finite at the Fermi level [Figs.~\ref{fig:pyro:dos}(B1)-(B4)] and 
the gap does not open until $U \sim 6$ [Fig.~\ref{fig:pyro:list}(a)].
The extrapolated values to $N \to \infty$ are summarized in Fig.~\ref{fig:pyro:list}(c)~\cite{LevelSpacing}. 

In summary, the above results reveal peculiar features 
in the ice-rule model at half filling, in sharp contrast to the random model: 
(i) The charge gap opens at a small $U_c$ compared to the bandwidth, 
(ii) DOS shows a 1D form in the large-$U$ charge-ice insulator, 
(iii) $\sigma(\omega)$ exhibits a rapid change for $U$ in the low-energy part and 
a suppression of a sharp spike in the large $U$ region, 
(iv) $w$ %$n_{\text{LEW}}$
shows a characteristic nonmonotonic behavior as a function of $U$, and 
(v) IPR suggests that a metal-insulator transition occurs at $U=U_c$, where the gap opens.
All these features are characteristic of the locally-correlated ice-rule potentials, which are not seen in the random systems.

%
%=============================================================================%
% SUBSECTION: Checkerboard Lattice                                            %
\subsection{\label{subsec:cboard}
Checkerboard lattice
}

\begin{figure}
  \begin{center}
     \includegraphics[width=3.33in]{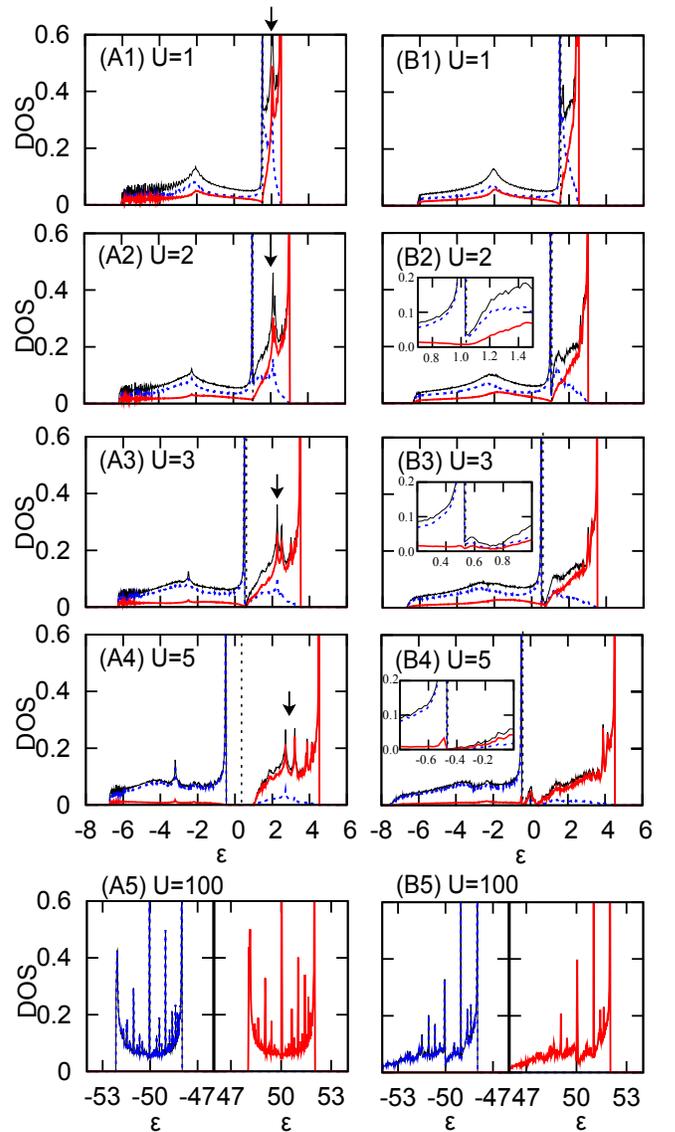}
  \end{center}
  \caption{
  (color online).
  DOS of itinerant fermions for the checkerboard lattice models with the ice-rule constraint (A1)-(A5)
  and with the random potential (B1)-(B5).  
  The insets of (B2), (B3), and (B4) show the enlarged figures of the main panels in the vicinity of the Fermi level at half filling.
  The symbols are common to those used in Fig.~\ref{fig:pyro:dos}.
  }
  \label{fig:cboard:dos}
\end{figure}

Figure~\ref{fig:cboard:dos} shows DOS for the checkerboard lattice models.
The left column (A1)-(A5) shows the numerical results for the ice-rule model, 
while the right column (B1)-(B5) for the random configuration. 
The symbols are common to those in Fig.~\ref{fig:pyro:dos}. 
The results are obtained for $3^2$ superlattices of $4\times 28^2$ sites.

DOS for the checkerboard lattice at $U=0$ consists of a dispersive band and a flat band,
similar to the pyrochlore lattice [Fig.~\ref{fig:lattice}(g)].
The dispersive band is equivalent to that of noninteracting
tight-binding model on the square lattice with an energy shift of $-2$,
and the flat band gives $\delta$-functional peak at $\varepsilon = 2$. 
The Fermi level at half filling is located just below the flat band;
the dispersive band is fully occupied while the flat band remains empty.
The system is metallic, in contrast to the pyrochlore case in which the system is semimetallic.

\begin{figure}
  \begin{center}
  \includegraphics[width=2.80in]{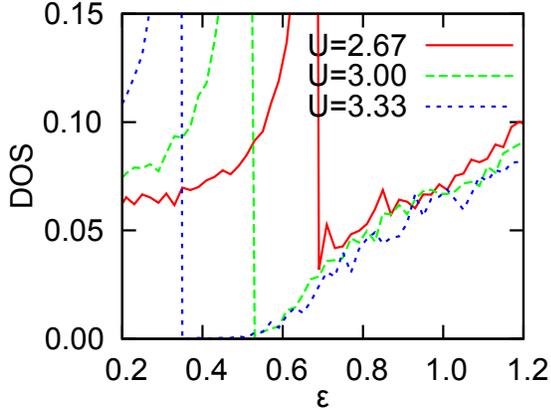}
  \end{center}
  \caption{
  (color online).
  DOS for the ice-rule model on the checkerboard lattice in the vicinity of the Fermi level at half filling.
  }
  \label{fig:cboard:gap}
\end{figure}

\begin{figure}
  \begin{center}
     \includegraphics[width=2.67in]{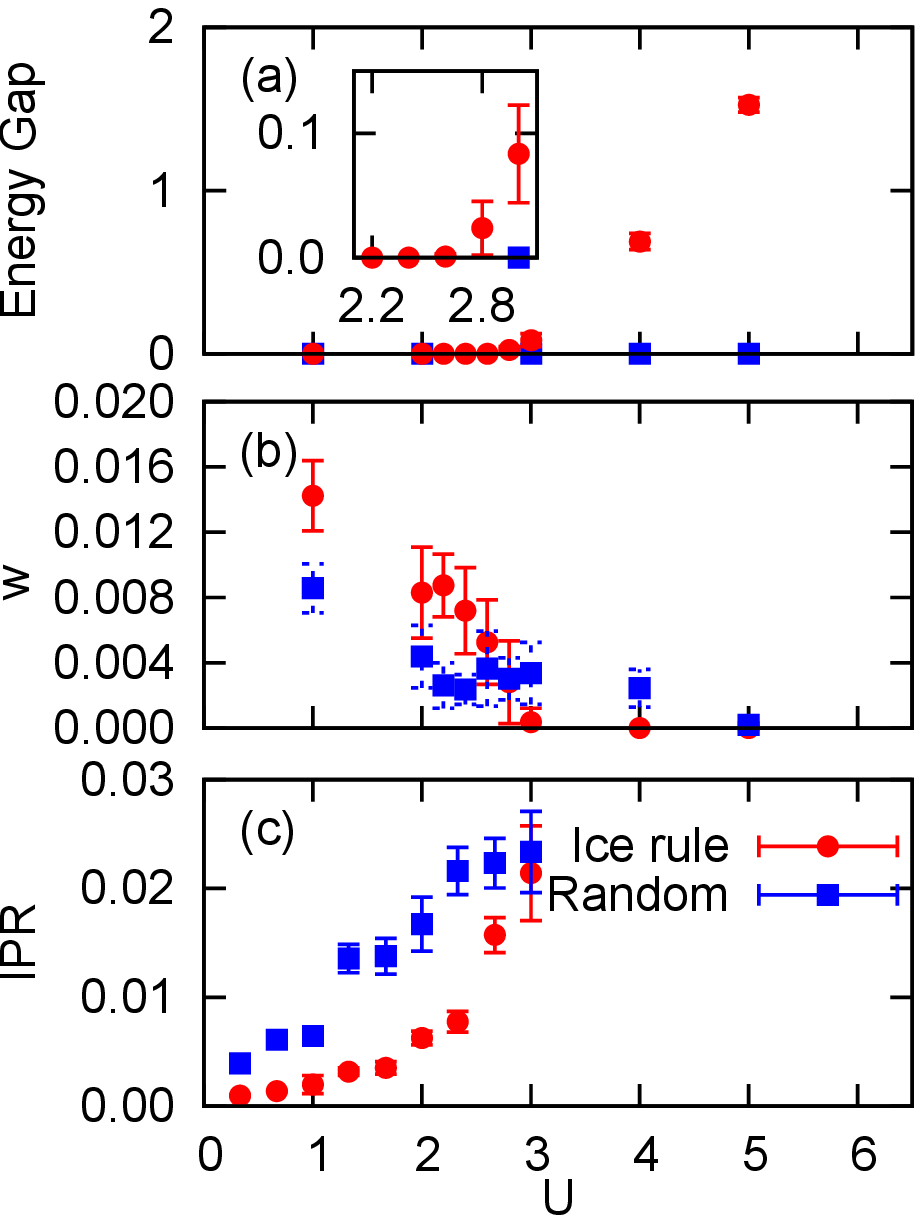}
  \end{center}
  \caption{
  (color online).
  $U$ dependence of (a) the energy gap, (b) the low-energy weight
  $w$%$n_{\text{LEW}}$
  , and
  (c) IPR for the checkerboard lattice models at half filling.
  $w$ %$n_{\text{LEW}}$
  is calculated by Eq.~(\ref{eq:neff}) with taking the cutoff $\omega_0 = 0.095$, and
  IPR is the extraporated value to $N\rightarrow \infty$. 
  The inset of (a) shows an enlarged view in the vicinity of $U_c = 2.7(2)$.
  }
  \label{fig:cboard:list}
\end{figure}

Despite of the difference in the dimensionality of lattice structures,
the evolution of DOS with increasing $U$ shows many common features with the pyrochlore lattice.
The flat band is perturbed to be broadened by $U$ and splits into two divergences, 
as shown in Figs.~\ref{fig:cboard:dos}(A1) and \ref{fig:cboard:dos}(B1). 
A cusp structure appears in between the two divergences for the ice-rule case 
as indicated by arrows in Figs.~\ref{fig:cboard:dos}(A1)-(A4). 
The Fermi level at half filling is pinned at the lower divergence. 
The energy gap opens at a smaller $U$ in the ice-rule case than in the random case 
[Figs.~\ref{fig:cboard:dos}(A3), (A4), (B3), and (B4); Fig.~\ref{fig:cboard:gap}; Fig.~\ref{fig:cboard:list}(a)]: 
The critical value of $U$ is estimated as $U_c = 2.7(2)$, 
which is markedly smaller than the bandwidth [see also Fig.~\ref{fig:cboard:list}(a)].
DOS at $U \rightarrow \infty$ approaches a similar form to 1D tight-binding model, 
as shown in Fig.~\ref{fig:cboard:dos}(A5). 
All these features are commonly seen in the pyrochlore lattice model as well as in the tetrahedron Husimi cactus
in Sec.~\ref{subsec:pyro}. 
These features illuminate the effect of the local ice-rule configuration characteristic of the corner-sharing tetrahedra,
working irrespective of global features such as lattice structures and dimensions of the system. 

\begin{figure}
  \begin{center}
    \includegraphics[width=3.33in]{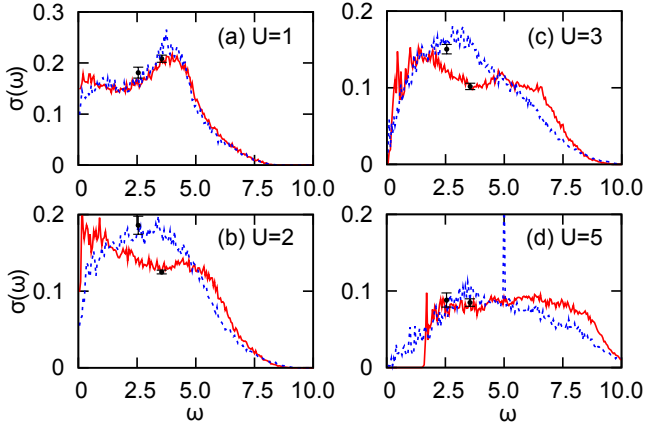}
  \end{center}
  \caption{
  (color online).
  Optical conductivity for the checkerboard lattice models at half filling. 
  The symbols are the same as in Fig.~\ref{fig:pyro:sigma}. 
  }
  \label{fig:cboard:sigma}
\end{figure}

Similar features to the pyrochlore lattice are also observed in the optical conductivity.
Figure~\ref{fig:cboard:sigma} shows the optical conductivity for the checkerboard lattice 
at half filling measured in the $x$ direction.
The symbols are common to those in Fig.~\ref{fig:pyro:sigma},
and the calculations were done on $3^2$ superlattices of $4\times 10^2$ sites.
The overall evolution of $\sigma(\omega)$ with increasing $U$ resembles that of the pyrochlore models; 
In the random case, a dip slowly develops at $\omega \sim 0$ and it remains as a pseudogap
with increasing $U$, while a gap opens in the ice-rule case corresponding to the gap opening of DOS.

However, when one carefully looks at the low-energy part of $\sigma(\omega)$,
there is a difference compared to the pyrochlore case.
Figure~\ref{fig:cboard:list}(b) shows the low-energy weight for the checkerboard lattice models.
$W$ for the ice-rule model shows a monotonic decrease with increasing $U$,
which is qualitatively different from the nonmonotonic behavior seen in the pyrochlore case
[Fig.~\ref{fig:pyro:list}(b)].  This is presumed to be owing to the  difference in DOS at $U=0$. 
As is described above, in the pyrochlore case, a semimetallic gap exists at the Fermi level of half filling,
on the other hand, the checkerboard lattice is metallic with finite DOS at the Fermi level. 
Hence, at $U=0$, $\sigma(\omega)$ is zero in the former case, but divergent in the latter.
The different behavior of $\sigma(\omega)$ in the small $U$ region 
presumably comes from this difference. 

\begin{figure}
  \begin{center}
     \includegraphics[width=3.33in]{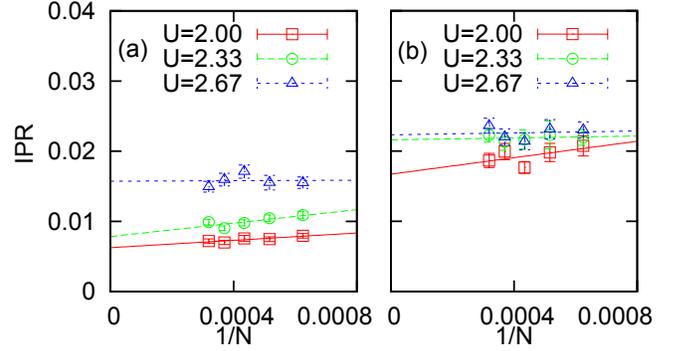}
  \end{center}
  \caption{
  (color online).
  IPR for the half-filled state in the checkerboard lattice models with
  (a) the ice-rule and (b) random potential. 
  }
  
  \label{fig:cboard:ipr}
\end{figure}

Further difference between the pyrochlore and checkerboard models is seen in IPR.
Figure~\ref{fig:cboard:ipr} shows IPR for the ice-rule [Fig.~\ref{fig:cboard:ipr}(a)] and
random case [Fig.~\ref{fig:cboard:ipr}(b)] with varying $U$.
Unlike the pyrochlore case, the extrapolated values of IPR for both ice-rule and random cases 
appear to remain finite at $U=2$,
which is definitely smaller than $U_c = 2.7(2)$.  This suggests that the wave functions 
near the Fermi level 
are localized in spite that DOS does not show a gap; namely, 
the system is Anderson insulator for both ice-rule and random cases.
This is in sharp contrast to the pyrochlore case,
where the ice-rule model appears to remain metallic until the gap opening at $U=U_c$.

To summarize, the checkerboard lattice show several features common to the pyrochlore lattice, 
in particular, in the evolution of DOS with increasing $U$.
On the other hand, transport properties exhibit differences, 
suggesting the qualitatively different nature of the metal-insulator transition 
in the ice-rule models between the pyrochlore and checkerboard cases. 
These similarities and differences will be discussed further in Sec.~\ref{sec:discussion}. 

%
%=============================================================================%
% SUBSECTION: Kagome Lattice                                                  %
%
\subsection{\label{subsec:kagome}
Kagome lattice
}

Next, we move to the kagome-ice model introduced in Sec.~\ref{sec:lattices}.
In this model, we focus on the $2/3$-filling case of itinerant fermions instead of half filling,
since we are interested in the transition to charge-ice insulating state 
which is expected to occur at 2/3 filling by confinement of fermions to $-U/2$ sites. 
We also make a comparison with the random model, in which $-U/2$ ($+U/2$) potential sites are randomly
distributed with keeping the ratio of the total number of sites at $2:1$.

\begin{figure*}
  \begin{center}
    \includegraphics[width=4.27in]{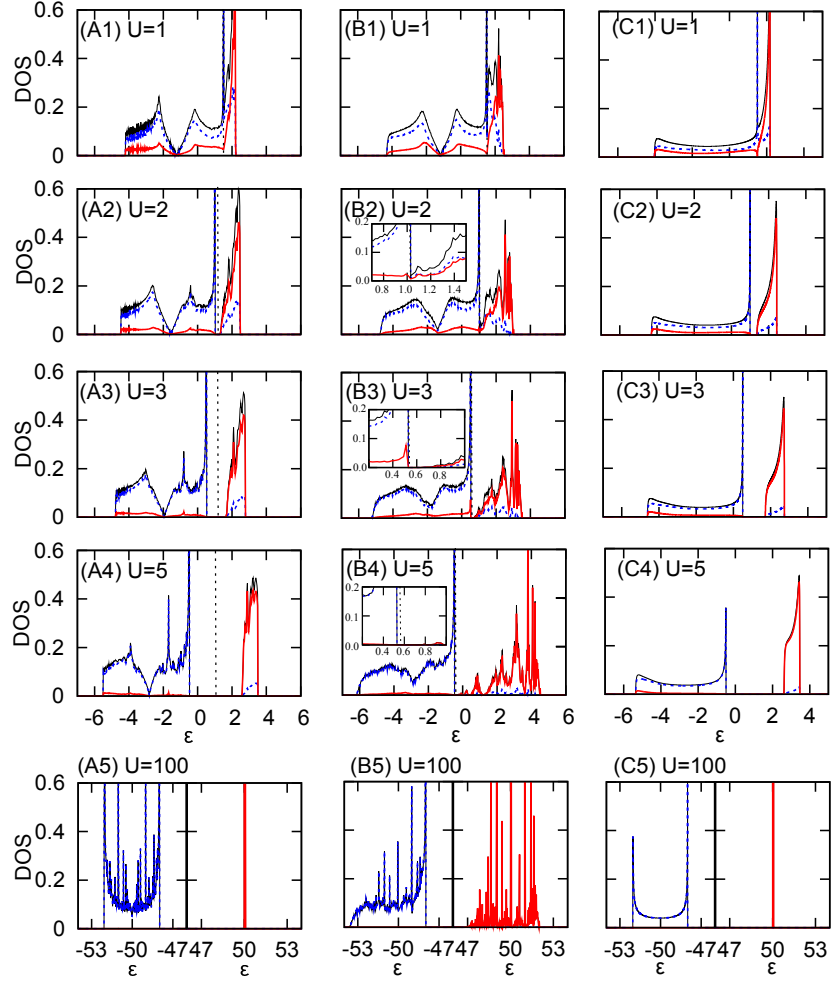}
  \end{center}
  \caption{
  (color online).
  DOS of itinerant fermions for the kagome models 
  with the kagome ice-rule constraint (A1)-(A5) and 
  with the random potential (B1)-(B5); 
  DOS for the triangle Husimi cactus model with the kagome ice-rule constraint (C1)-(C5).
  The insets of (B2), (B3), and (B4) show the enlarged figures of the main panels in the vicinity of the Fermi level at 2/3 filling.
  The symbols are common to those in Figs.~\ref{fig:pyro:dos} and \ref{fig:cboard:dos}.
  }
  \label{fig:kagome:dos}
\end{figure*}

Figure~\ref{fig:kagome:dos} shows DOS for the kagome models. 
Each column shows the numerical results for the kagome ice-rule case, 
the corresponding random case, and the exact solution for the triangle Husimi cactus model,
respectively from left to right. Each row shows the results for varying $U$.
The calculations for the kagome models are conducted for $3^2$ superlattices of $3\times 36^2$ sites.
The results for the triangle Husimi cactus model are obtained by Eqs.~(\ref{fullDyson}) and (\ref{g_triangle12}).

DOS for the kagome lattice model at $U=0$ consists of two 
dispersive bands which form the continuum spectrum at $-4\leq \varepsilon \leq 2$
and a flat band at $\varepsilon=2$ [Fig.~\ref{fig:lattice}(h)].
The dispersive bands are equivalent to those of honeycomb lattice, and a semimetallic dip exists at $\varepsilon= -1$, 
where the lower two bands linearly cross by forming the Dirac points.
At $2/3$ filling, the Fermi level is located just below the flat band and
the system is metallic similar to the checkerboard lattice model. 

\begin{figure}
  \begin{center}
  \includegraphics[width=2.80in]{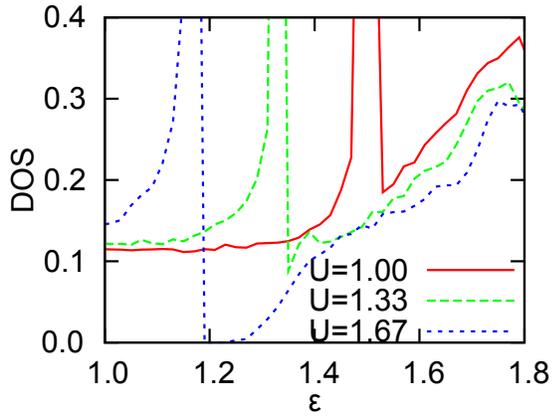}
  \end{center}
  \caption{
  (color online).
  DOS for the ice-rule model on the kagome lattice in the vicinity of the Fermi level at 2/3 filling.
  }
  \label{fig:kagome:gap}
\end{figure}

\begin{figure}
  \begin{center}
     \includegraphics[width=2.67in]{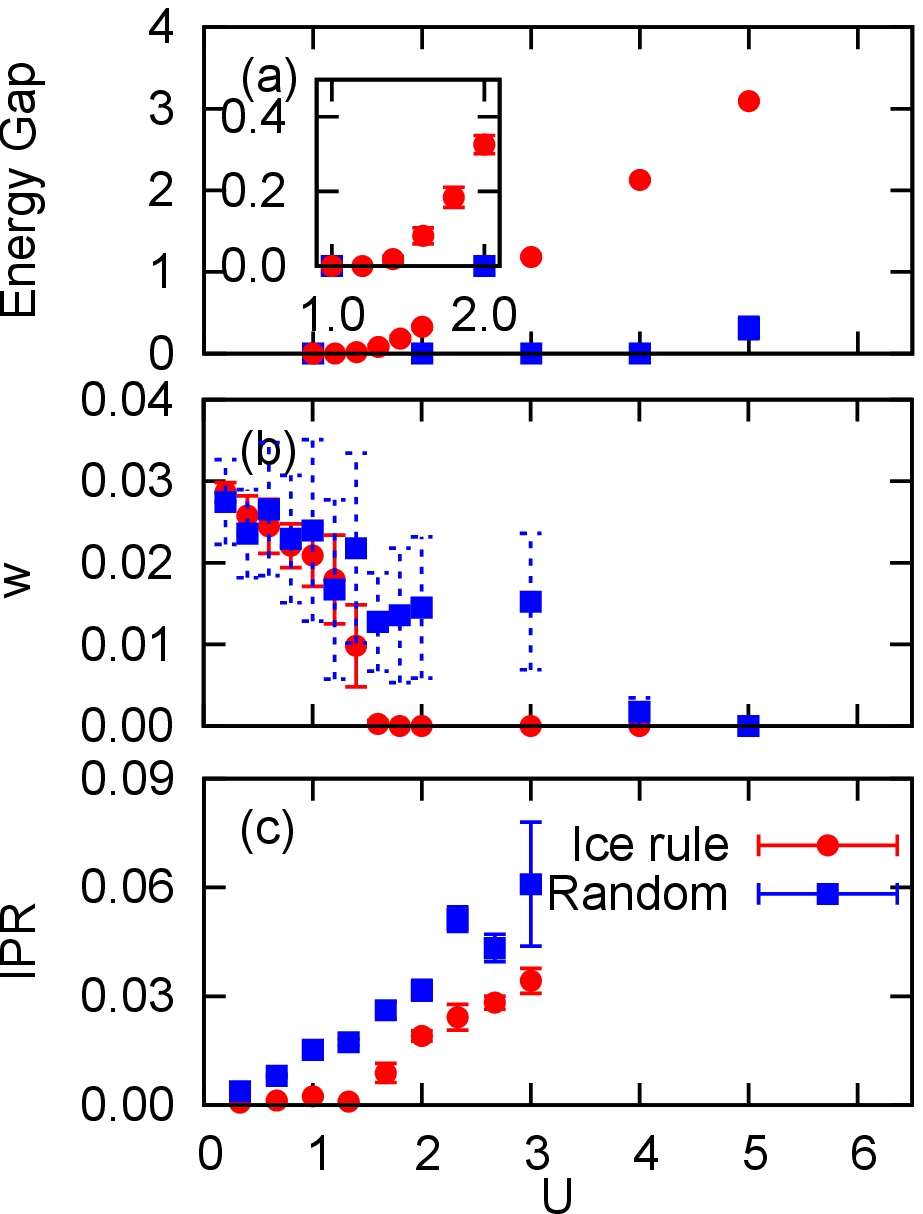}
  \end{center}
  \caption{
  (color online).
  $U$ dependence of (a) the energy gap, (b) the 
  low-energy weight $w$, and (c) IPR for the kagome models at 2/3 filling. 
  $w$ is calculated by Eq.~(\ref{eq:neff}) with taking the cutoff
  $\omega_0 = 0.095$, and
  IPR is the extraporated value to $N\rightarrow \infty$. 
  The inset of (a) shows an enlarged view in the vicinity of $U_c = 1.4(2)$ 
  }
  \label{fig:kagome:list}
\end{figure}

With switching on $U$, the flat band is broadened and the lower edge shows a divergence: 
The Fermi level at 2/3 filling is located at the divergence 
[Figs.~\ref{fig:kagome:gap}(A1) and \ref{fig:kagome:gap}(B1)].
With further increasing $U$, in the kagome ice-rule case, the gap starts to open at a 
fairly small $U_c = 1.4(2)$ compared to the bandwidth 
[Fig.~\ref{fig:kagome:gap}(A2); Fig.~\ref{fig:kagome:gap}; Fig.~\ref{fig:kagome:list}(a)].
On the contrary, for the random case, 
the gap does not open until $U \sim 4-5$, while a quasigap feature develops for $U \gtrsim 3$ 
[Figs.~\ref{fig:kagome:gap}(B2)-(B4); Fig.~\ref{fig:kagome:list}(a)].
These behaviors are qualitatively similar to the pyrochlore and checherboard lattice models.

On the other hand, the evolution of DOS toward large $U$ limit is different from the preceding
two models due to difference in the geometrical unit of the lattice. 
In the kagome ice-rule model, the bandwidth of the upper band shrinks with
increasing $U$ for $U>U_c$ [Figs.~\ref{fig:kagome:dos}(A2)-(A4)],
and finally,  becomes a $\delta$-functional peak, as shown in Fig.~\ref{fig:kagome:dos}(A5). 
Correspondingly, a cusp-like structure appearing between two divergences is obscure
[Figs.~\ref{fig:kagome:dos}(A1)-(A4)], in contrast to the pyrochlore and checkerboard cases.
On the other hand, the lower band approaches a 1D-like form.
These behaviors are explained by considering the strong coupling limit:
In contrast to the pyrochlore and checkerboard cases,
the sites with $U_i=+U/2$ are disconnected from each other in the limit of $U\rightarrow\infty$ in the kagome-ice model, 
while the $-U/2$ sites form 1D loops [see Fig.~\ref{fig:lattice}(c)]. 
Consequently, localized states at each isolated $+U/2$ sites contribute to the $\delta$-functional divergence, 
and the states for 1D loops with $-U/2$ potential give the 1D-like DOS. 

The evolution of DOS for the kagome-ice rule case is well captured 
by the triangle Husimi cactus model, as shown in Figs.~\ref{fig:kagome:dos}(C1)-(C5). 
In this cactus, the critical value of $U$ for the gap opening is $U_c=1$, 
which is a half of $U_c=2$ for the tetrahedron Husimi cactus.

Another peculiar feature of DOS for the kagome model is the evolution of
characteristic structures in the dispersive bands, such as the semimetallic dip 
between the two dispersive bands (at $\varepsilon = -1$ for $U=0$),
and the van-Hove singularity (at $\varepsilon = -2$ for $U=0$). 
In the kagome ice-rule case, these features remain to be clearly seen even for $U=5$, 
while they are smeared out already for $U=3$ in the random case.
Note that similar tendency can be seen in the pyrochlore and checkerboard cases.

\begin{figure}
  \begin{center}
    \includegraphics[width=3.33in]{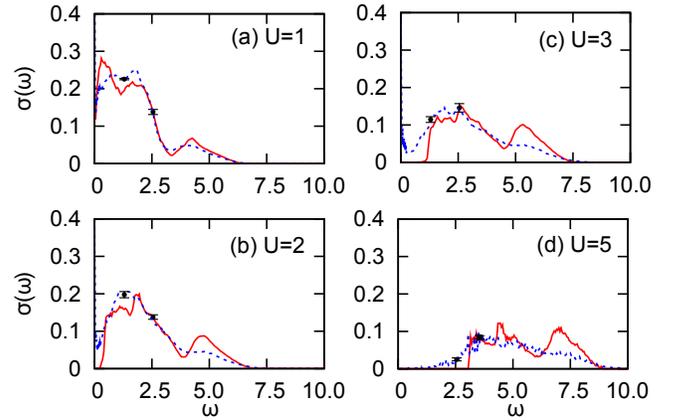}
  \end{center}
  \caption{
  (color online).
  Optical conductivity for the kagome models at 2/3 filling. 
  The symbols are the same as in Figs.~\ref{fig:pyro:sigma} and \ref{fig:cboard:sigma}.
  }
  \label{fig:kagome:sigma}
\end{figure}

Next we examine transport properties of this system.
The optical conductivity at 2/3 filling behaves in a qualitatively similar manner to that for the checkerboard model, 
monotonically decreasing with increasing $U$ as shown in Fig.~\ref{fig:kagome:sigma}.
The calculation is done for $4^2$ superlattices of $3\times 12^2$ sites.
A gap in the low-$\omega$ region rapidly develops in the kagome-ice case corresponding to the gap opening in DOS,
whereas $\sigma(\omega)$ changes rather slowly in the random case. 
The low-energy weight $w$ 
is also shown in Fig.~\ref{fig:kagome:list}(b).
$w$ decreases monotonically with increasing $U$ 
and goes to zero at $U \simeq U_c$ in the kagome ice-rule case, 
similarly to the checkerboard model in Fig.~\ref{fig:cboard:list}(b). 

\begin{figure}
  \begin{center}
     \includegraphics[width=3.33in]{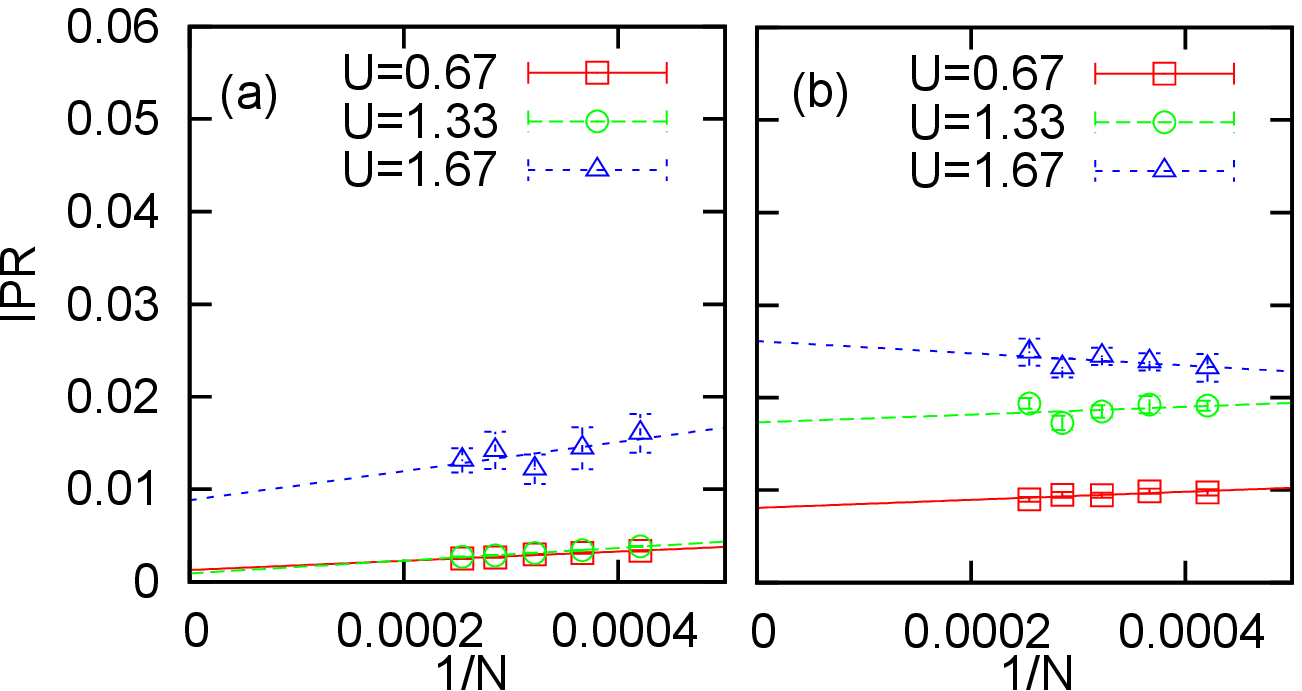}
  \end{center}
  \caption{
  (color online).
  IPR for the 2/3-filled state in the kagome models with 
  (a) the kagome-ice rule and (b) the random cases. 
  }
  \label{fig:kagome:ipr}
\end{figure}

The result of IPR for this system is shown in Fig.~\ref{fig:kagome:ipr}. 
With sufficiently small $U$,
the results for the kagome-ice model appear to show the wave functions to remain extended,
while they appear to be localized with very small $U$ in the random case.
The extrapolated values of IPR shown in Fig.~\ref{fig:kagome:list}(c) indicates this
difference more clearly: The extrapolated values of IPR remain to be almost zero for $U<U_c$ in
the ice-rule case, while they are finite even at $U=1$ in the random case. 
These behaviors will be discussed further in comparison with 
the pyrochlore and checkerboard cases in Sec.~\ref{sec:discussion}. 

As a consequence, our calculations for the kagome-ice model suggest a transition from metal to gapped charge-ice insulator
without any clear indication of Anderson localization. 
On the other hand, in the random case, the Anderson insulator exists before the gapped insulating state. 
The results will be discussed in the next section, in comparison with those for the other lattice models. 

\section{\label{sec:discussion}
Discussions
}

All the results in Sec.~\ref{sec:result} clearly elucidate 
peculiar electronic properties under the influence of ice-rule correlation, 
which are distinctive from those in the completely random case. 
Furthermore, the comparative studies among different lattice structures revealed common features and differences.
We discuss the universality and diversity in the following.

First, we discuss several universal features of DOS common to the three models we have investigated through this paper.
All the models exhibit a considerably small critical value of $U$ for gap opening, $U_c$.
The values of $U_c$ are estimated as $2.3(2)$, $2.7(2)$, and $1.4(2)$ for the pyrochlore,
checkerboard, and kagome models, respectively.
These values are markedly smaller than the bandwidth at $U=0$, $W$;
$W=8$ for the pyrochlore and checkerboard lattices, and $W=6$ for the kagome lattice.
On the other hand, for the random case, the energy gap evolves around $U\sim W$ for all the three 
models, reflecting the competition between the local binary potential and kinetic energy.
This contrastive behavior of $U_c$ indicates that the ice-rule short-range correlation 
helps to develop a gapped insulating state, compared to the uncorrelated random case.  

Second, DOS shows a common feature in the large $U$ regime. 
For all lattice structures, at $U\rightarrow\infty$,
itinerant fermions are excluded from $U_i=+U/2$ sites and localized at $U_i=-U/2$ site 
forming ice-rule charge distribution complementary to that of localized particles. 
We call this insulating state the ``charge ice" insulator. 
The distribution of fermions is a set of 1D loops made of $-U/2$ potential sites. 
Reflecting the specific configurations, DOS approaches a universal symmetric form given by the
1D tight-binding form or the $\delta$ function.
This behavior of DOS in the large $U$ limit is also qualitatively different from those for random models:
In the random cases, DOS becomes highly asymmetric and broadened, irrespective of the lattice structure.

In addition, we note that the characteristic features in the noninteracting band structure are
well preserved for the ice-rule cases compared with the random cases.
For example, the van-Hove singularities and the semimetallic dip are retained up to fairly large $U$,
while such structures are rapidly smeared out for the random case.

It is noteworthy that these universal features of DOS are well captured by the exact solutions for the Husimi cactus models. 
The cactus models also show relatively small $U_c$ for the transition to charge ice insulator;
$U_c=2$ ($U_c=1$) compared to the bare bandwidth $W=8$ ($W=6$) for the tetrahedron (triangle) Husimi cactus.
Moreover,  at $U\rightarrow\infty$, the cactus models show qualitatively similar
DOS as the original lattice models; a 1D tight-binding form (plus a $\delta$ function)
for the tetrahedron (triangle) Husimi cactus. 
The agreement not only indicates that the cactus models give good references to the original
lattice models but also confirms that these features of DOS are universal among the ice-rule systems. 

In contrast to DOS, transport properties are dependent on the lattice structures. 
For the pyrochlore lattice, the optical conductivity $\sigma(\omega)$,
the low-energy weight $w$,
and IPR consistently show that the system becomes insulating at $U=U_c$. 
This suggests that the metal-insulator transition 
occurs simultaneously with the transition to the charge ice insulator,
and that it is not driven by the Anderson localization before the gap opening. 
Interestingly, the low-energy weight indicates a nonmonotonic change with increasing $U$, 
presumably owing to the semimetallic nature at $U=0$ [Fig.~\ref{fig:pyro:list}(b)]. 

The results for the checkerboard lattice look different.
While the energy gap opens at $U_c=2.7(2)$ for this model,
IPR indicates that the system is insulating already for $U\lesssim 2$,
as shown in Fig.~\ref{fig:cboard:list}. 
This appears to contradict with a finite $w$ %$n_{\text{LEW}}$
remaining up to $U\simeq U_c$; 
however, it is hard to estimate the true low-energy contribution 
because of the sharp dip structure of $\sigma(\omega)$ at $\omega \sim 0$ %, 
[Fig.~\ref{fig:cboard:sigma}]. 
Thus the results indicate that the Anderson insulator region appears in a wide region of $U$ before
the energy gap opens at $U_c \sim 2.7(2)$. 
The difference of transport properties between the pyrochlore and checkerboard models 
may be attributed to the difference in dimensionality of the system.
It is well known that the Anderson localization is relevant for lower-dimensional systems.

From the viewpoint of dimensionality,
the kagome model would share qualitative aspects of the transport properties with the checkerboard model. 
Our numerical results, however, do not clearly show the Anderson localization (Fig.~\ref{fig:kagome:list}). 
A possible reason for the apparent absence of the Anderson localization is that
it occurs in a narrow range of $U$ beyond our numerical resolution. 
The smaller $U_c\sim 1.4(2)$ makes it harder to observe the Anderson transition if any. 

Recently, the Anderson localization in the systems with flat bands was
analysed~\cite{Chalker2010}. It was shown that a random potential broadens the
flat bands, and the states originating from the flat bands become critical, 
neither Anderson localized nor spatially extended. 
The analysis was done for, typically, a Gaussian distribution of the potential, 
which smears out the flat band divergence. 
In our case, the potential distribution is binary 
originating in the configuration of localized particles, and hence, 
the divergences remain for finite amplitude of the potential.  
The problems are related with each other, but the situations are different. 
Our work focuses on the metal-insulator transition in the systems 
under the ice-rule correlated potentials; 
the results cast a new issue on the localization problems 
from the characteristic spatial correlation of potential configurations.

\section{\label{sec:summary}
Summary
}

In summary, we have studied the extended Falicov-Kimball model on the pyrochlore, checkerboard, and kagome lattices. 
With exact diagonalization of Hamiltonian and approximation of the statistical average by the arithmetic mean,
we obtained the density of states, optical conductivity, and IPR of the models.
Through the analysis, we have clarified how the local ice-rule constraint affects the global electronic structure, 
reflecting the peculiar nature of ice-rule manifold.

The ice-rule local constraint gives rise to several universal features in the density of states.
The distinctive features are summarized as follows.
(i) The energy gap opens at a much smaller $U$ compared to the bandwidth. 
(ii) In the large $U$ limit, 
the density of states becomes a similar form to that of one-dimensional tight-binding model (or a $\delta$ functional peak).
(iii) A cusp-like structure appears as a precursor of the band-edge divergence in the large $U$ limit.
These features emerge from the interaction with localized particles under the ice rule,
and are not observed in the coupling to random potentials.
Furthermore, they are insensitive to the details of lattice structure, 
and can be captured by the exactly-solvable cactus models.
The universal behaviors in the characteristic evolution of the density of states 
can be considered as a hallmark of transition to the charge-ice insulating state.

In contrast, transport properties depend on the detailed lattice structure, 
such as dimensionality and local geometrical unit.
For the pyrochlore lattice model, the optical conductivity and the inverse participation ratio consistently
indicate that a metal-insulator transition takes place accompanied with the gap opening,
in sharp contrast to the Anderson localization for the random case. 
The low-energy weight of the optical conductivity shows a nonmonotonic behavior as a function of $U$ in the metallic region.
On the contrary, transport properties for the checkerboard lattice model look different.
Contrastive to the pyrochlore case, the low-energy weight decreases monotonically with respect to $U$. 
Moreover, the results of the inverse participation ratio and energy gap suggest 
Anderson localization in the region $U < U_c$.
The contrasting behaviors are ascribed to the difference of dimensions of the systems. 
Yet different behaviors appear for the kagome lattice models which have a different geometrical unit, a triangle: 
our results show no clear evidence of the Anderson localization for $U < U_c$.
This is presumably 
due to the small $U_c$ or the limited precision of our calculations. 

Our results clearly indicate the significance of local correlations on the global electronic structure and transport properties. 
Such local correlations should be responsible for the peculiar properties 
observed in several pyrochlore compounds~\cite{Taguchi2001,Nakatsuji2006,Nakatsuji2010}, where 
itinerant electrons interact with localized moments under the ice-rule type constraint.
For understanding these interesting properties, 
it will be crucial to take into account the characteristic spatial correlations 
emerging from the local constraint.

\begin{acknowledgements}

The authors thank S. Nishino, K. Penc, H. Shinaoka, and Y. Yamaji for fruitful discussions.
This work was supported by KAKENHI (Nos. 17071003, 19052008, 21740242, and 21340090), 
Global COE Program ``the Physical Sciences Frontier," 
and by the Next Generation Super Computing Project, Nanoscience Program, MEXT, Japan.
\end{acknowledgements}

\appendix
\section{\label{subsec:groundstate}
Lifting of the ground state degeneracy 
}
%The numerical results presented in the main text are all obtained by 
%the arithmetic mean over ice-rule configurations of localized particles.
In the main text, in replacing the statistical average by the arithmetic mean,
we assumed that the Boltzmann weight for all teh ice-rule configuration is equivalent.
However,  in the extended Falicov-Kimball model, Eq.~(\ref{eq:hamiltonian_fk}), 
the coupling to itinerant fermions may lift the degeneracy of ice-rule manifold and select
a unique ground state or a submanifold. 
Hence, to properly investigate the thermal properties of this model,
we need to take account of the energy levels in the Boltzmann weight in Eq.~(\ref{observable}). 
Nevertheless, we expect our results to be plausible at low temperatures 
as we will show below in this section.
This is due to the fact that lifting of degeneracy is very small in the entire range of $U$,
specifically, in the order of $10^{-4}t$.

\begin{figure}[t] 
  \hspace{12mm}
  \begin{center}
   \includegraphics[width=3.33in]{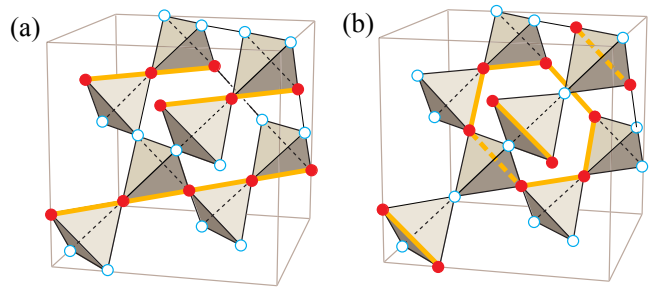}
  \caption{(color online).
  Schematic pictures of 
  (a) A-type stripe and (b) hexagonal ordered states.
  The sites with filled (open) circles represent $-U/2$ ($+U/2$) sites.
  The highlighted bonds in (a) shows stripes of $-U/2$ and (b) represents hexagons of $-U/2$ potential sites.
  See the text for details.
  }
  \label{fig:gse:order}
  \end{center}
\end{figure}

To investigate the effect of coupling to itinerant fermions, 
we first consider the problem by the perturbation in $t/U$ in the large $U$ limit.
For the pyrochlore lattice model, the lowest-order contribution to the energy difference 
comes from the ring-exchange-type hopping process on a hexagon embedded 
in the pyrochlore lattice structure (see Fig.~\ref{fig:gse:order}).
The contribution is in the sixth-order of $t/U$,
and depends on the number of localized particles in the hexagon. 
The energy becomes the lowest for a hexagon with three localized particles. 
The relative energy to a fictitious reference system in which all hexagons have three localized
particles is given by 
\begin{eqnarray}
\Delta E= \frac{t^6}{U^5}\left\{ 12(n_0+n_6) + 10(n_1+n_5) + 20(n_2+n_4) \right\},\nonumber\\ \label{eq:pyro:deltaE}
\end{eqnarray}
where $n_m$ ($0\leq m\leq 6$) gives the number of hexagons in which $m$ localized particles exist.
Notes that the energy depends only on the number of localized particules in each exagon, not on their
configurations within each hexagon.
Since the contribution starts from the order of $t^6/U^5$, 
the energy difference among different ice-rule configurations remains 
very small in the large-$U$ charge-ice insulating regime. 

To see how the energy difference develops as $U$ decreases beyond the perturbation regime, 
we numerically evaluate the energy difference in the entire range of $U$. 
Since the energy difference is very small and 
becomes comparable to the numerical resolution of our calculations 
when we perform the arithmetic mean over the randomly generated samples, 
we try to estimate a typical difference by considering 
two specific periodic configurations with very different number of $n_m$ hexagons. 
One is the A-type stripe order shown in Fig.~\ref{fig:gse:order}(a),
in which localized particles exist only on [110] chains. 
The other is the hexagonal ordered state shown in Fig.~\ref{fig:gse:order}(b),
in which one-third of hexagons in all the [111] kagome layers are fully occupied by
localized particles. 
These two states are largely different in the numbers of different-type hexagons as shown in
Table~\ref{tab:gse:dist}, thus,
the ground state energy is expected to be substantially different in the scheme of perturbation theory. 
We compare the ground-state energies of itinerant fermions for these two systems by numerical diagonalization method.
The calculation was done on $16^3$ superlattices 
of $4\times 3^3$ sites.

\begin{table}[t]
  \begin{center}
  \begin{tabular}{|c|c|c|c|c|c|c|c|}
   \hline
      & \ $n_0$ \ & \ $n_1$ \ & \ $n_2$ \ & \ $n_3$ \ & \ $n_4$ \ & \ $n_5$ \ & \ $n_6$ \ \\
   \hline
    stripe    &  0    & 0     & 2     & 0     & 2     & 0     & 0     \\
    hexagonal &  0    & 0     & 4/3   & 2     & 1/3   & 0     & 1/3   \\
   \hline
  \end{tabular}
  \caption{
  Distributions of hexagons with $m$ localized particles for A-type stripe and hexagonal ordered states.
  The values are normalized to the number of each hexagon in a cubic unit cell shown 
  in Fig.~\ref{fig:gse:order}. 
  }
  \label{tab:gse:dist}
  \end{center}
\end{table}

\begin{table}[t]
  \begin{center}
  \begin{tabular}{|c|c|c|c|}
   \hline
      \ $U/t$ \ & \ \ \ stripe \ \ \ & \ hexagonal \ & \ difference \ \\
   \hline
     2 & -3.20546 & -3.20474  & $ -7.2 \times 10^{-4}$ \\
     5 & -2.67592 & -2.67658  & $  6.7 \times 10^{-4}$ \\
     10 & -2.37914 & -2.37924 & $  0.9 \times 10^{-4}$ \\
   \hline
  \end{tabular}
  \caption{Ground state energy for the A-type stripe and hexagonal ordered states per site 
  and their difference.
  The data are for $16^3$ superlattices 
  of $4\times3^3$ sites.
  }
  \label{tab:gse:result}
  \end{center}
\end{table}

The results are shown in Table~\ref{tab:gse:result}.
From the numerical results in a broader range of $U$ including the metallic region,
we find that the energy variance remains surprisingly small, typically in the order of $10^{-3}t$ or less. 

The weak lifting of the ice-rule manifold implies that the manifold is preserved down to very low 
$T$ even in the ex-FK model:
The Boltzmann weights for different ice-rule configurations are virtually the same for $T/t \gg 10^{-3}$. 
On the other hand, the ice-rule constraint of the localized particles is expected to be well retained for
$T \lesssim U$.
In such wide $T$ region, thermal average might be well approximated by the arithmetic mean as performed in
the present study. 
Such situation was considered also in our previous study~\cite{Udagawa2010}.

The situation is similar for the kagome lattice models. 
In this case also, the smallest loop is the six-site hexagon, and hence, 
the discussions above can be applied straightforwardly. 

For the checkerboard lattice, the same approach gives fourth-order perturbation to be the lowest order
\begin{eqnarray}
\Delta E= \frac{t^4}{U^3}\left\{ 2(n_0+n_4) + 6 n_2 \right\}, \label{eq:cboard:deltaE}
\end{eqnarray}
since the smallest loop is on the four-site plaquette [see Fig.~\ref{fig:lattice}(b)].
This gives larger energy shift compared to the pyrochlore and kagome lattice cases, and hence,
is expected to be more relevant 
on lifting the ground state degeneracy. 
For this reason, some orderings may be expected for the checkerboard lattice models at 
low $T$. 
Nevertheless, in this paper, we limit ourselves to treating this model with the arithmetic mean and leave the
possibility of ordering for future problem.


\begin{thebibliography}{99}
\bibitem{autocorr}      $\tau$ is estimated from the autocorrelation function of the density of localized
                        particles, $\langle n_i^f(t) n_i^f(0)\rangle$. We also confirm that the distribution
			function of loop length converges to a universal form for $t > \tau$.
\bibitem{IPR}           We also compute the single state IPR for an unoccupied state which has the numerically lowest
                        energy above the flat band, and confirm that the results give similar conclusions.
\bibitem{LevelSpacing}  We also calculated the distribution function of the nearest level spacings, which is another
                        standard measure of the localization. The results show consistent tendency with those of IPR, 
                        whereas the convergence of the distribution function is not satisfactory 
                        because of small number of available eigenstates.
\end{thebibliography}
\end{document}